\documentclass[aip,jcp,amsmath,amssymb,reprint,superscriptaddress]{revtex4-2}
\usepackage{uvt-paper}
\begin{document}

\title{fix uvt and fix pimd/uvt: A Unified LAMMPS Framework for Constant-Potential Constant-Temperature Molecular Dynamics}

\author{Li Fu}
\affiliation{School of Materials Science and Engineering, Peking University, Beijing 100871, People's Republic of China}
\author{Yifan Li}
\email{yifanl0716@gmail.com}
\affiliation{Department of Chemistry, Princeton University, Princeton, NJ 08544, USA}
\author{Shenzhen Xu}
\email{xushenzhen@pku.edu.cn}
\affiliation{School of Materials Science and Engineering, Peking University, Beijing 100871, People's Republic of China}

\begin{abstract}

Accurate simulations of electrochemical interfaces require the simultaneous treatment of constant-potential conditions, nuclear quantum effects, and sufficient configurational sampling. Integrating these capabilities within a general and efficient molecular dynamics (MD) framework remains challenging. In this work, we implement \texttt{fix uvt} and \texttt{fix pimd/uvt} in LAMMPS for constant-potential classical MD and path integral molecular dynamics (PIMD), respectively. We organize the class hierarchy to reuse LAMMPS's existing Nosé--Hoover chain thermostat routines and share nuclear propagation routines across PIMD integrators. A common interface connects these integrators to models that provide the electron-number derivative of the potential energy. We present three examples with accompanying input commands to guide users through constant-potential classical MD, thermostatted PIMD, and constant-potential PIMD, covering analytical models, liquid water, and electrochemical interfaces described by machine learning potentials. This work provides practical tools and guidance for large-scale constant-potential simulations incorporating nuclear quantum effects.

\end{abstract}

\maketitle

\section{Introduction}
\label{sec:1}

Atomistic simulations provide a powerful approach for understanding electrochemical processes at electrode--electrolyte interfaces, particularly proton-coupled electron transfer (PCET) reactions.\cite{Huynh2007ProtonCoupled,HammesSchiffer2001Theoretical,Warburton2022Theoretical,Levell2024Emerging} However, a reliable description of these processes requires the simultaneous treatment of several key physical factors. First, the electronic state should respond to an external electron reservoir under constant-potential conditions, rather than being constrained to a fixed electron number or charge state.\cite{Bonnet2012FirstPrinciples,Sundararaman2017Grand,Xia2023Grand,Sun2026Electrochemical} Second, PCET reactions involve light hydrogen nuclei and can therefore exhibit significant nuclear quantum effects (NQEs), even at room temperature.\cite{Fu2026ElectrochemistryEnhanced,Sun2025Probing,Ceriotti2016Nuclear,Markland2018Nuclear} In addition, realistic electrode--electrolyte interfaces exhibit substantial structural and solvent fluctuations, requiring sufficient configurational sampling of the heterogeneous interfacial environment.\cite{Sun2025Probing,Sakong2020Water,Sarabia2024Exploring} Meeting these requirements simultaneously remains computationally demanding, particularly when realistic interfaces and reactive configurations must be sampled over molecular dynamics (MD) timescales.\cite{Fu2026ElectrochemistryEnhanced,Sun2025Probing,Tian2025Electrochemical,Heenen2020Solvation}

Conventional atomistic simulations of electrochemical interfaces are often performed under fixed-charge or fixed-electron-number conditions, rather than under an electronically grand canonical (GC) condition in which the charge state can adjust in response to an imposed electrode potential.\cite{Levell2024Emerging,Hormann2024Converging} To address this limitation, a variety of constant-potential approaches have been developed, ranging from variable-charge electrode models\cite{Siepmann1995Influence,Reed2007Electrochemical,Hu2025Observation} to GC electronic-structure methods\cite{Sundararaman2017Grand,Xia2023Grand,Lozovoi2001Ab} and extended-variable potentiostat approaches\cite{Bonnet2012FirstPrinciples,Ikeshoji2017FirstPrinciples,Ikeshoji2017Toward}. Constant-potential MD schemes have consequently been formulated for both empirical force fields and first-principles simulations; in many implementations, however, the nuclear degrees of freedom (DOFs) are still propagated classically, neglecting NQEs.\cite{Levell2024Emerging,Bonnet2012FirstPrinciples,Reed2007Electrochemical}

Path integral molecular dynamics (PIMD) provides an established framework for incorporating NQEs into atomistic simulations by representing quantum nuclei as ring polymers composed of multiple beads.\cite{Tuckerman1993Efficient,Feynman1948SpaceTime,Feynman2010Quantum} Although converged PIMD simulations typically require tens to hundreds of beads, making them computationally expensive, machine learning interatomic potentials (MLIPs) pave the way for efficient PIMD simulations with first-principles accuracy.\cite{Buxton2017Accelerated,Markland2008An,Bocus2023Nuclear,Li2026Fix,Li2022Using} A recent extension to i-PI, a versatile and lightweight Python-based framework that has become a de facto standard for PIMD,\cite{Kapil2019IPI,Litman2024IPI} supports constant-potential MD through interfaces to electronic structure programs and provides a route toward future integration with MLIPs.\cite{Zhang2026A} To enable efficient large-scale PIMD simulations with MLIPs, we previously implemented PIMD in the $NVT$ and $NpT$ ensembles using a Langevin thermostat within the widely used LAMMPS framework.\cite{Li2026Fix,Thompson2022LAMMPS} However, this implementation lacks constant-potential support, motivating its extension to constant-potential classical MD and PIMD for interfacial electrochemical simulations.

In our previous work, we established a GC path integral formulation for constant-potential quantum dynamics and applied it to PCET rate calculations.\cite{Fu2026ElectrochemistryEnhanced} The corresponding implementation, however, was developed for that specific application and was not integrated as a reusable component within the existing LAMMPS PIMD infrastructure. In this work, we develop a LAMMPS-native framework that integrates thermostatted-potentiostatted classical molecular dynamics (TP-Classical MD) and thermostatted-potentiostatted path integral molecular dynamics (TP-PIMD) within a common simulation infrastructure. We provide \texttt{fix uvt} as a constant-potential integrator for classical MD. We refactor the existing \texttt{fix pimd/langevin} and \texttt{fix pimd/nvt} implementations to share nuclear propagation routines through the newly added \texttt{fix pimd/nve}, improving code maintainability. We then reuse the Nosé--Hoover chain (NHC) thermostat implementation underlying \texttt{fix nvt}, together with the shared PIMD propagation routines, to implement \texttt{fix pimd/uvt} for TP-PIMD simulations. To handle the electronic response, we provide a general interface that decouples the propagation algorithm from the specific method used to evaluate the electron-number derivative. We demonstrate the use of the framework and validate its implementation through TP-Classical MD simulations of an analytical model and MLIP-based electrochemical simulations using both classical MD and PIMD. Together, these capabilities provide a general and user-friendly LAMMPS workflow for constant-potential simulations incorporating NQEs.

\section{Theory}
\label{sec:2}

In this section, we briefly recapitulate the theoretical formulations of TP-Classical MD and TP-PIMD, including their extended Hamiltonians and equations of motion, which form the basis of the implementation described in Sec.~\ref{sec:2-2}.

\subsection{Classical MD in the \texorpdfstring{\(\mu VT\)}{muVT} Ensemble}
\label{sec:2-1-1}

TP-Classical MD targets the GC ensemble under simultaneous control of temperature \(T\) and electrochemical potential \(\mu_{\mathrm{e}}\).\cite{Bonnet2012FirstPrinciples,Sundararaman2017Grand,Xia2023Grand} For a system in equilibrium with an electron reservoir, the grand partition function can be expressed as

\begin{equation}
\label{eq:1}
\itXi(\beta,\mu_{\mathrm{e}}) = \sum_{N_{\mathrm{e}}}^{}{\itLambda\int \mathrm{d}\bm{R}\exp\lbrack - \beta\itPhi\left( \bm{R},N_{\mathrm{e}} \right)\rbrack},
\end{equation}

where the grand potential is defined as

\begin{equation}
\label{eq:2}
\itPhi\left( \bm{R},N_{\mathrm{e}} \right) = U\left( \bm{R},N_{\mathrm{e}} \right) - \mu_{\mathrm{e}}N_{\mathrm{e}}.
\end{equation}
Here, \(\beta = \left( k_{\text{B}}T \right)^{- 1}\), \(\bm{R} ={\{\bm{r}_{i}\}}_{i = 1}^{N}\) denotes the particle positions, where \(N\) is the number of particles. \(U\left( \bm{R},N_{\mathrm{e}} \right)\) is the potential energy as a function of the particle positions and the system's total electron number \(N_{\mathrm{e}}\), and \(\itLambda\) arises from integrating out the particle momenta. In the extended-variable formulation used below, \(N_{\mathrm{e}}\) is treated as a continuous dynamical variable,\cite{Bonnet2012FirstPrinciples,Fu2026ElectrochemistryEnhanced} providing a continuous extension of the discrete GC electron-number sum for dynamical sampling.

To sample the target GC distribution, TP-Classical MD combines a potentiostat for electrochemical-potential control with a Nosé--Hoover thermostat for temperature control.\cite{Bonnet2012FirstPrinciples,Fu2026ElectrochemistryEnhanced,Martyna1992NosHoover} The extended Hamiltonian is written as

\begin{equation}
\label{eq:3}
\begin{aligned}
H_{\mathrm{cl}}^{\mathrm{ext}} ={}& \sum_{i = 1}^{N}\frac{\bm{p}_{i}^{2}}{2m_{i}} + U\left( \bm{R},N_{\mathrm{e}} \right)+\frac{p_{\eta}^{2}}{2Q} + gk_{\mathrm{B}}T\eta + \frac{p_{N_{\mathrm{e}}}^{2}}{2m_{N_{\mathrm{e}}}} - \mu_{\mathrm{e}}N_{\mathrm{e}}.
\end{aligned}
\end{equation}
where \(\bm{p}_{i}\) and \(m_{i}\) denote the momentum and mass of particle \(i\). A conjugate momentum \(p_{N_{\mathrm{e}}}\) and an associated mass \(m_{N_{\mathrm{e}}}\) are introduced for \(N_{\mathrm{e}}\). The variables \(\eta\) and \(p_{\eta}\) are the generalized coordinate and momentum associated with the Nosé--Hoover thermostat, and \(Q\) is the associated mass. The parameter \(g = f + 1\), where \(f\) is the number of thermostatted particle momentum DOFs and the additional unity accounts for the momentum DOF associated with \(N_{\mathrm{e}}\). Following the standard non-canonical transformation to physical variables,\cite{Bonnet2012FirstPrinciples,Fu2026ElectrochemistryEnhanced} the equations of motion take the form

\begin{equation}
\label{eq:4}
\left\{ \begin{array}{r@{}l}
{\dot{\bm{r}}}_{i} &{}= \frac{\bm{p}_{i}}{m_{i}}, \\
{\dot{\bm{p}}}_{i} &{}= - \frac{\partial U\left( \bm{R},N_{\mathrm{e}} \right)}{\partial\bm{r}_{i}} - \frac{p_{\eta}}{Q}\bm{p}_{i}, \\
\dot{\eta} &{}= \frac{p_{\eta}}{Q}, \\
{\dot{p}}_{\eta} &{}= \sum_{i = 1}^{N}\frac{\bm{p}_{i}^{2}}{m_{i}} - gk_{\mathrm{B}}T + \frac{p_{N_{\mathrm{e}}}^{2}}{m_{N_{\mathrm{e}}}}, \\
{\dot{N}}_{\mathrm{e}} &{}= \frac{p_{N_{\mathrm{e}}}}{m_{N_{\mathrm{e}}}}, \\
{\dot{p}}_{N_{\mathrm{e}}} &{}= - \frac{\partial U\left( \bm{R},N_{\mathrm{e}} \right)}{\partial N_{\mathrm{e}}} + \mu_{\mathrm{e}} - \frac{p_{\eta}}{Q}p_{N_{\mathrm{e}}}.
\end{array} \right.
\end{equation}

The negative derivative of the potential energy with respect to the total electron number is denoted by \(W_{\mathrm{e}}\) (work function) and referred to as the electronic response,

\begin{equation}
\label{eq:5}
W_{\mathrm{e}}\left( \bm{R},N_{\mathrm{e}} \right) \equiv - \frac{\partial U\left( \bm{R},N_{\mathrm{e}} \right)}{\partial N_{\mathrm{e}}}.
\end{equation}

The resulting dynamics conserve the extended Hamiltonian \(H_{\mathrm{cl}}^{\mathrm{ext}}\) and generate the GC distribution for the physical variables,

\begin{equation}
\label{eq:6}
\begin{aligned}
\rho_{\mathrm{cl}}\left(\bm{R},\bm{P},N_{\mathrm{e}}\right)
\propto\exp\Biggl[-\beta\Biggl(\sum_{i = 1}^{N}\frac{\bm{p}_{i}^{2}}{2m_i} + U\left(\bm{R},N_{\mathrm{e}}\right)-\mu_{\mathrm{e}}N_{\mathrm{e}}\Biggr)\Biggr],
\end{aligned}
\end{equation}
where $\bm{P}=\{\bm{p}_i\}_{i=1}^{N}$ denotes the set of particle momenta.

Accordingly, the constant-potential condition at equilibrium is $\left\langle -W_{\mathrm{e}}\left(\bm{R},N_{\mathrm{e}}\right)\right\rangle=\mu_{\mathrm{e}}$, where $\langle\ldots\rangle$ denotes an ensemble average over the GC distribution. This identity follows by integrating the electron-number derivative of Eq.~\eqref{eq:6} with vanishing boundary terms, as derived in Sec.~S6 of the Supplementary Material (SM). The non-canonical transformation and corresponding phase-space derivation have been detailed previously.\cite{Bonnet2012FirstPrinciples,Fu2026ElectrochemistryEnhanced} In practical simulations, the single Nosé--Hoover thermostat is generalized to a Nosé--Hoover chain.\cite{Martyna1992NosHoover}

\subsection{PIMD in the \texorpdfstring{\(\mu VT\)}{muVT} Ensemble}
\label{sec:2-1-2}

TP-PIMD extends the constant-potential formulation to quantum nuclear sampling by replacing the classical particle DOFs with a \(P\)-bead ring polymer representation.\cite{Tuckerman1993Efficient,Feynman1948SpaceTime,Feynman2010Quantum} The corresponding ring polymer Hamiltonian in the Cartesian coordinate is

\begin{equation}
\label{eq:8}
\begin{aligned}
H_{\mathrm{RP}} ={}& \sum_{b=1}^{P}\sum_{i=1}^{N}\Biggl[\frac{\left(\bm{p}_i^{(b)}\right)^2}{2m_i} +\frac{1}{2}m_i\omega_P^2\left(\bm{r}_i^{(b)}-\bm{r}_i^{(b+1)}\right)^2\Biggr] \\
&+\frac{1}{P}\sum_{b=1}^{P}U\left(\bm{R}^{(b)},N_{\mathrm{e}}\right),
\end{aligned}
\end{equation}
where \(P\) is the number of beads and \(b = 1,\ 2,\ldots,P\) denotes the bead index, \(\bm{r}_{i}^{(P + 1)} = \bm{r}_{i}^{(1)}\), and \(\omega_{P} = \sqrt{P}\text{/}(\beta\hslash)\) in the physical-temperature representation. Here, \(\bm{R}^{(b)} = {\{\bm{r}_{i}^{(b)}\}}_{i = 1}^{N}\) and \(\bm{P}^{(b)} = {\{\bm{p}_{i}^{(b)}\}}_{i = 1}^{N}\) denote the particle positions and momenta of bead \(b\). The same continuous electron-number variable \(N_{\mathrm{e}}\) is shared by all beads, such that each bead potential is evaluated as \(U\left( \bm{R}^{(b)},N_{\mathrm{e}} \right)\). The requirement of a shared \(N_{\mathrm{e}}\) among all beads originates from the orthogonality of electronic states with different electron numbers in Fock space.\cite{Fu2026ElectrochemistryEnhanced,Sun2025Probing}

For propagation, the ring polymer coordinates and momenta are transformed from the Cartesian coordinate \(\{\bm{R}^{(b)},\bm{P}^{(b)}\}\) to the standard normal mode representation \(\left\{ {\widetilde{\bm{R}}}^{(k)},{\widetilde{\bm{P}}}^{(k)} \right\}\), where \(k = 0,\ 1,\ldots,P - 1\) labels the normal modes.\cite{Tuckerman1993Efficient} The transformed ring polymer Hamiltonian is

\begin{equation}
\label{eq:9}
\begin{aligned}
\widetilde{H}_{\mathrm{RP}} ={}& \sum_{k=0}^{P-1}\sum_{i=1}^{N}\Biggl[\frac{\left(\widetilde{\bm{p}}_i^{(k)}\right)^2}{2m_i} +\frac{1}{2}m_i\left(\widetilde{\omega}^{(k)}\right)^2\left(\widetilde{\bm{r}}_i^{(k)}\right)^2\Biggr] \\
&+\frac{1}{P}\sum_{b=1}^{P}U\left(\bm{R}^{(b)},N_{\mathrm{e}}\right),
\end{aligned}
\end{equation}
where \({\widetilde{\omega}}^{(k)}\) denotes the frequency of the \(k\)-th normal mode. The physical potential energy is evaluated using bead coordinates reconstructed from the inverse normal mode transformation. The extended ring polymer Hamiltonian in the normal mode representation becomes

\begin{equation}
\label{eq:10}
\begin{aligned}
\widetilde{H}_{\textnormal{TP-PIMD}}^{\mathrm{ext}} ={}& \widetilde{H}_{\mathrm{RP}} +\sum_{k=0}^{P-1}\left[\frac{p_{\eta^{(k)}}^2}{2Q^{(k)}}+g_k k_{\mathrm{B}}T\eta^{(k)}\right] +\frac{p_{N_{\mathrm{e}}}^2}{2m_{N_{\mathrm{e}}}}-\mu_{\mathrm{e}}N_{\mathrm{e}},
\end{aligned}
\end{equation}
where \(\eta^{(k)}\), \(p_{\eta^{(k)}}\), and \(Q^{(k)}\) denote the Nosé--Hoover thermostat coordinate, conjugate momentum, and mass parameter associated with the \(k\)-th normal mode, respectively. The parameter \(g_{k} = f_{k} + 1\text{/}P\), where \(f_{k}\) is the number of particle momentum DOFs associated with the \(k\)-th normal mode, and the \(1/P\) term accounts for the shared electronic momentum \(p_{N_{\mathrm{e}}}\).

The equations of motion corresponding to the TP-PIMD extended Hamiltonian are

\begin{equation}
\label{eq:11}
\left\{\begin{array}{r@{}l}
\dot{\widetilde{\bm{r}}}_i^{(k)} &{}= \dfrac{\widetilde{\bm{p}}_i^{(k)}}{m_i}, \\[3pt]
\dot{\widetilde{\bm{p}}}_i^{(k)} &{}= -m_i\left(\widetilde{\omega}^{(k)}\right)^2\widetilde{\bm{r}}_i^{(k)}+\widetilde{\bm{F}}_i^{(k)}-\dfrac{p_{\eta^{(k)}}}{Q^{(k)}}\widetilde{\bm{p}}_i^{(k)}, \\[3pt]
\dot{\eta}^{(k)} &{}= \dfrac{p_{\eta^{(k)}}}{Q^{(k)}}, \\[3pt]
\dot{p}_{\eta^{(k)}} &{}= \displaystyle\sum_{i=1}^{N}\frac{\left(\widetilde{\bm{p}}_i^{(k)}\right)^2}{m_i}-g_k k_{\mathrm{B}}T+\dfrac{1}{P}\dfrac{p_{N_{\mathrm{e}}}^2}{m_{N_{\mathrm{e}}}}, \\[3pt]
\dot{N}_{\mathrm{e}} &{}= \dfrac{p_{N_{\mathrm{e}}}}{m_{N_{\mathrm{e}}}}, \\[3pt]
\dot{p}_{N_{\mathrm{e}}} &{}= -\displaystyle\frac{1}{P}\sum_{b=1}^{P}\frac{\partial U\left(\bm{R}^{(b)},N_{\mathrm{e}}\right)}{\partial N_{\mathrm{e}}}+\mu_{\mathrm{e}}-\displaystyle\frac{1}{P}\sum_{k=0}^{P-1}\frac{p_{\eta^{(k)}}}{Q^{(k)}}p_{N_{\mathrm{e}}},
\end{array}\right.
\end{equation}
where \({\widetilde{\bm{F}}}_{i}^{(k)}\) denotes the normal mode transform of the bead-space effective forces \(\bm{F}_{i}^{(b)}= - \frac{1}{P}\partial U\left( \bm{R}^{(b)},N_{\mathrm{e}} \right)/\partial\bm{R}_{i}^{(b)}\), which are evaluated independently for each bead. The electronic response of each bead is defined as

\begin{equation}
\label{eq:12}
W_{\mathrm{e}}^{(b)} \equiv - \frac{\partial U\left( \bm{R}^{(b)},N_{\mathrm{e}} \right)}{\partial N_{\mathrm{e}}},
\end{equation}
and the response driving the shared electron-number DOF is the bead-averaged quantity
\begin{equation}
\label{eq:bead-averaged-response}
\overline{W}_{\mathrm{e}}=\frac{1}{P}\sum_{b=1}^{P}W_{\mathrm{e}}^{(b)}.
\end{equation}

The resulting dynamics conserve the extended Hamiltonian \({\widetilde{H}}_{\textnormal{TP-PIMD}}^{\mathrm{ext}}\) and generate the quantum GC distribution of the ring polymer system,

\begin{equation}
\label{eq:14}
\begin{aligned}
&\rho_{\textnormal{TP-PIMD}}\left(\left\{\widetilde{\bm{R}}^{(k)},\widetilde{\bm{P}}^{(k)}\right\},N_{\mathrm{e}}\right) \\
\propto&\exp\Biggl[-\beta\Biggl(\widetilde{H}_{\mathrm{RP}}\left(\left\{\widetilde{\bm{R}}^{(k)},\widetilde{\bm{P}}^{(k)}\right\},N_{\mathrm{e}}\right)-\mu_{\mathrm{e}}N_{\mathrm{e}}\Biggr)\Biggr].
\end{aligned}
\end{equation}

Accordingly, the constant-potential condition at equilibrium is

\begin{equation}
\label{eq:15}
\left\langle - {\overline{W}}_{\mathrm{e}} \right\rangle = \mu_{\mathrm{e}}.
\end{equation}

In practical simulations, the single Nosé--Hoover thermostat is generalized to a Nosé--Hoover chain.\cite{Martyna1992NosHoover}

\section{Software Implementation and Usage}
\label{sec:2-2}
\begin{figure*}[t]
\centering
\includegraphics[width=0.8\linewidth]{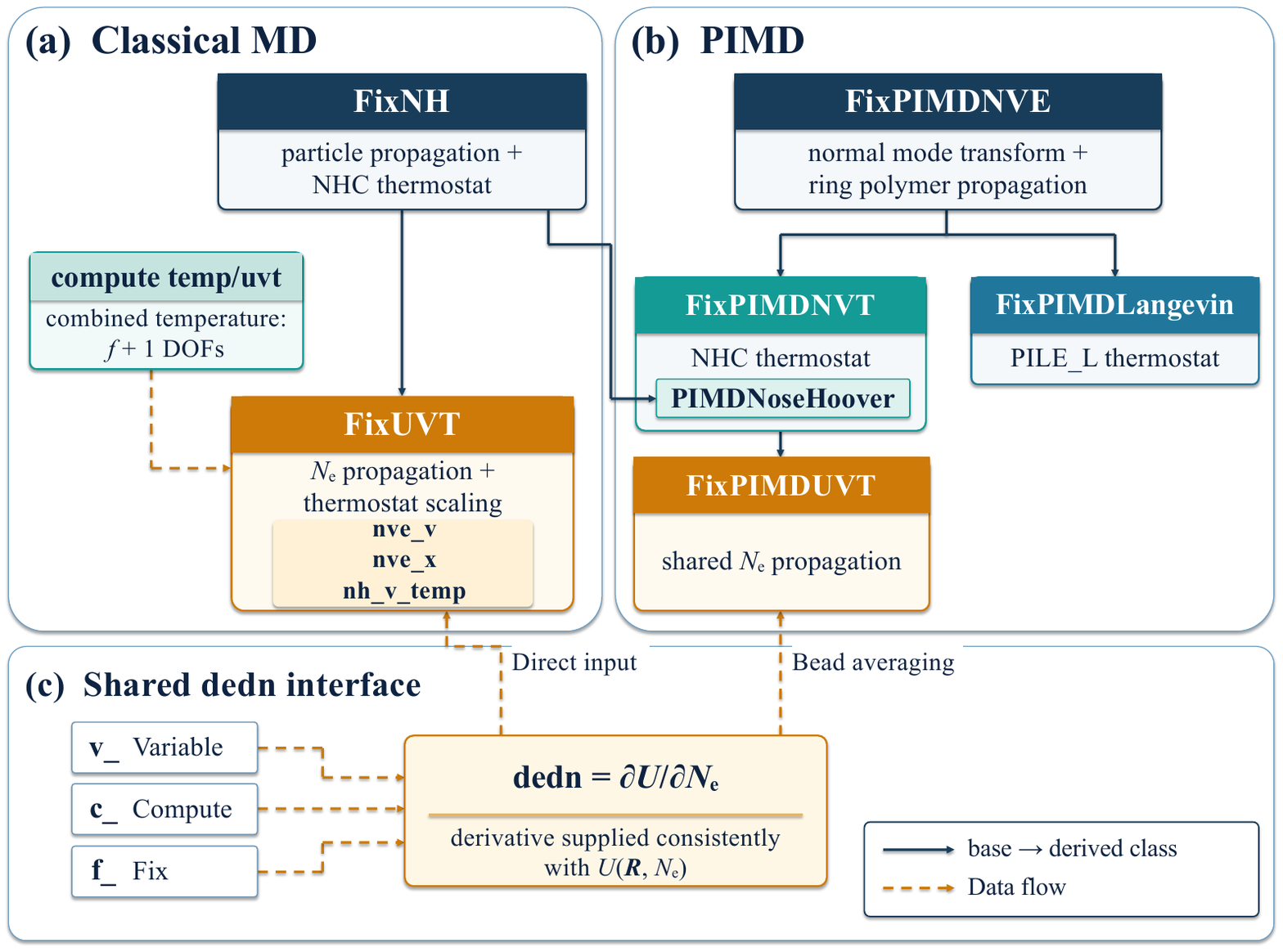}
\caption{Class hierarchies and shared electronic-response interface for the LAMMPS implementation. (a) \texttt{FixUVT} derives from \texttt{FixNH}, extending particle propagation and Nos\'e--Hoover chain thermostatting to the \(N_{\mathrm{e}}\) DOF. The \texttt{compute temp/uvt} style supplies the combined temperature and total number of DOFs. (b) \texttt{FixPIMDNVT} and \texttt{FixPIMDLangevin} derive from \texttt{FixPIMDNVE} and provide NHC thermostatting and the local version of the path integral Langevin equation (PILE\_L) thermostat, respectively. The internal \texttt{PIMDNoseHoover} adapter in \texttt{FixPIMDNVT} derives from \texttt{FixNH} to reuse its NHC routines; \texttt{FixPIMDUVT} derives from \texttt{FixPIMDNVT} and propagates the shared \(N_{\mathrm{e}}\). (c) The common \texttt{dedn} interface accepts \(\partial U/\partial N_{\mathrm{e}}\) from an equal-style variable (\texttt{v\_}), a compute (\texttt{c\_}), or a fix (\texttt{f\_}). The derivative enters \texttt{FixUVT} directly and is averaged over beads in \texttt{FixPIMDUVT}. Solid arrows indicate inheritance from parent to derived class; dashed arrows indicate data flow.}
\label{fig:1}
\end{figure*}

In this section, we describe the LAMMPS-native implementations of classical MD and PIMD in the constant potential ensemble and briefly introduce their usage. The implementation introduces two LAMMPS fix styles: \texttt{fix uvt} for TP-Classical MD and \texttt{fix pimd/uvt} for TP-PIMD.\cite{Thompson2022LAMMPS} We organize the class hierarchy so that the new integrators inherit from existing LAMMPS classes and reuse their routines wherever possible, minimizing code duplication and improving maintainability. FIG.~\ref{fig:1} summarizes this design for classical MD and PIMD, together with the electronic-response interface shared by the two fix styles.

\subsection{Implementation and Usage of TP-Classical MD}
For classical MD, we introduce two source files, \texttt{fix\_uvt.h} and \texttt{fix\_uvt.cpp}, which define the \texttt{FixUVT} class derived from LAMMPS's \texttt{FixNH} base class, as shown in FIG.~\ref{fig:1}(a). \texttt{FixUVT} extends \texttt{FixNH} by introducing the \(N_{\mathrm{e}}\) DOF, whose kinetic energy contributes to the total kinetic energy. An NHC thermostat acts on all \(g = f + 1\) DOFs, including the \(f\) particle DOFs and the additional \(N_{\mathrm{e}}\) DOF. To account for this additional contribution when evaluating the temperature, we implement the \texttt{compute temp/uvt} style, which calculates the instantaneous temperature as
\begin{equation}
    \label{eq:combined-temperature}
    T_{\mathrm{ins}} = \frac{1}{gk_{\mathrm{B}}}\left(\sum_{i=1}^{N}\frac{\bm{p}_{i}^{2}}{m_{i}} + \frac{p_{N_{\mathrm{e}}}^{2}}{m_{N_{\mathrm{e}}}}\right).
\end{equation}
Here, the two terms in parentheses are twice the particle kinetic energy and twice the kinetic energy associated with \(N_{\mathrm{e}}\), respectively. $T_{\mathrm{ins}}$ is then passed to the \texttt{FixNH} base class for thermostat propagation.

The source code of \texttt{FixNH} is reused to propagate the nuclear and thermostat DOFs. In \texttt{FixUVT}, we only add three new lines for propagating the $N_{\mathrm{e}}$ DOF: in the \texttt{nve\_v} function, we update $\dot{N}_{\mathrm{e}}$ according to the potentiostat force $-\frac{\partial U}{\partial N_{\mathrm{e}}}+\mu_{\mathrm{e}}$; in the \texttt{nve\_x} function, we update $N_{\mathrm{e}}$ using $\dot{N}_{\mathrm{e}}$; in the \texttt{nh\_v\_temp} function, we update $\dot{N}_{\mathrm{e}}$ according to the thermostat. All three functions first call the corresponding functions of \texttt{FixNH} to propagate nuclear and thermostat DOFs, and then update the $N_{\mathrm{e}}$ DOF.

The derivative $\partial U/\partial N_{\mathrm{e}}$ is supplied through the \texttt{dedn} keyword. The electronic-response interface is detailed in Subsection~\ref{sec:2-2-2}. The following input fragment illustrates the use of \texttt{fix uvt} at the temperature and electrochemical potential used for the Deep Potential (DP) interface example in Sec.~\ref{sec:pimd-dp}:

\begin{grayverb}[nobreak=false]
\begin{Verbatim}
compute dEdN all deepmd/fparam/dedn f_tp[1]
fix tp all uvt temp 300 300 0.05 mu -3 -3 &
    0.05 ne 0.648702 dedn c_dEdN
thermo_style custom step f_tp[1] temp f_tp[2]
\end{Verbatim}
\end{grayverb}

Here, \texttt{compute dEdN} evaluates $\partial U/\partial N_{\mathrm{e}}$ using the DP model, with \texttt{f\_tp[1]} referencing the excess electron number $N_{\mathrm{e}}^{\mathrm{extra}}$ provided by \texttt{fix tp} to the model, following the convention described in Sec.~\ref{sec:classical-dp}. The \texttt{fix tp} command propagates the system in the $\mu VT$ ensemble. Following the LAMMPS convention, \texttt{temp Tstart Tstop Tdamp} specifies the target temperatures at the start and end of the run and the thermostat damping time; similarly, \texttt{mu Mustart Mustop Mudamp} specifies the initial and final target electrochemical potentials and the potentiostat damping time. Both damping parameters have time units and determine the first thermostat mass $Q=gk_{\mathrm{B}}T_{\mathrm{target}}(\texttt{Tdamp})^2$ and the electron-number mass $m_{N_{\mathrm{e}}}=fk_{\mathrm{B}}T_{\mathrm{target}}(\texttt{Mudamp})^2$, respectively, with $g=f+1$. The \texttt{ne} keyword sets the initial electron number, and \texttt{dedn c\_dEdN} specifies the source of $\partial U/\partial N_{\mathrm{e}}$.
In this example, the \texttt{fix uvt} style reports $N_{\mathrm{e}}^{\mathrm{extra}}$ and the instantaneous temperature $T_{\mathrm{ins}}$ including the contribution from the $N_{\mathrm{e}}$ DOF through \texttt{f\_tp[1]} and \texttt{f\_tp[2]}, respectively. The \texttt{thermo\_style} command outputs $N_{\mathrm{e}}^{\mathrm{extra}}$, the instantaneous particle temperature, and $T_{\mathrm{ins}}$ defined in Eq.~\eqref{eq:combined-temperature}, which includes the contribution from the $N_{\mathrm{e}}$ DOF. Both temperatures fluctuate around the target temperature, but their values slightly differ.

\subsection{Implementation and Usage of TP-PIMD}
\label{sec:2-2-1}

The existing LAMMPS styles \texttt{fix pimd/nvt} and \texttt{fix pimd/langevin} were developed separately to support NHC and Langevin thermostatting, respectively, and maintain distinct nuclear propagation and communication routines.\cite{Li2026Fix} To introduce constant-potential PIMD while reusing these capabilities, we reorganize the code into the class hierarchy shown in FIG.~\ref{fig:1}(b). The common base class \texttt{FixPIMDNVE} provides nuclear propagation, normal mode transformations, and inter-bead communication, while derived classes add thermostatting, barostatting, or electronic propagation. This separation reduces code duplication and improves maintainability.

The \texttt{FixPIMDNVE} class propagates the ring polymer in the $NVE$ ensemble without thermostatting and should be called with \texttt{fix pimd/nve}. The following input command illustrates an $NVE$ PIMD simulation:
\begin{grayverb}[nobreak=false]
\begin{Verbatim}
fix 1 all pimd/nve method nmpimd temp 300
\end{Verbatim}
\end{grayverb}
Here, \texttt{temp} specifies the physical temperature $T$ used to define $\omega_P$ and does not activate a thermostat. Our PIMD implementation follows the convention used in i-PI,\cite{Kapil2019IPI,Litman2024IPI} defining the ring polymer frequency as $\omega_P=P/(\beta\hslash)$ and, when thermostatting is enabled, targeting the scaled temperature $PT$ internally. This convention differs from the physical-temperature formulation in Sec.~\ref{sec:2-1-2}, but users should still specify the physical target temperature $T$ in the input file; the implementation applies the scaling internally. The two formulations sample the same physical ensemble. Further details are given in Sec.~S1 of the SM and our previous \texttt{fix pimd/langevin} paper.\cite{Li2026Fix}

The \texttt{FixPIMDNVT} class derives from \texttt{FixPIMDNVE} and adds an NHC thermostat for each bead or normal mode.\cite{Martyna1992NosHoover,Tuckerman1993Efficient} We reuse the existing NHC implementation in LAMMPS's \texttt{FixNH} class, avoiding the need to reimplement the thermostat. To achieve this, we introduce a dedicated adapter class, \texttt{PIMDNoseHoover}, derived from \texttt{FixNH}, to access its \texttt{nhc\_temp\_integrate} and \texttt{nh\_v\_temp} functions for thermostat-chain propagation and velocity scaling, respectively. An instance of this adapter is an internal object of \texttt{FixPIMDNVT} and is not exposed to users. An example input command for an $NVT$ PIMD simulation using \texttt{fix pimd/nvt} is:
\begin{grayverb}[nobreak=false]
\begin{Verbatim}
fix 1 all pimd/nvt method nmpimd temp 300 &
    Tdamp 0.05 tchain 3
\end{Verbatim}
\end{grayverb}

The \texttt{FixPIMDLangevin} class also derives from \texttt{FixPIMDNVE} and supports the local version of the path integral Langevin equation (PILE\_L) thermostat.\cite{Ceriotti2010Efficient} In addition to $NVT$ simulations, it supports the $NpT$ ensemble, which is not available with \texttt{fix pimd/nvt}. Usage of \texttt{fix pimd/langevin} is detailed in Ref.~\citenum{Li2026Fix}. An example command for an $NVT$ simulation is:
\begin{grayverb}[nobreak=false]
\begin{Verbatim}
fix 1 all pimd/langevin ensemble nvt temp &
    300 thermostat PILE_L 1234 tau 0.05
\end{Verbatim}
\end{grayverb}

The constant-potential extension is implemented by \texttt{FixPIMDUVT}, which derives from \texttt{FixPIMDNVT}. It adds one electron-number coordinate $N_{\mathrm{e}}$ shared by all beads, together with its velocity and fictitious mass. Each bead is assigned to a separate LAMMPS partition,\cite{Thompson2022LAMMPS} with all ranks within the bead participating in the evaluation of $\partial U/\partial N_{\mathrm{e}}$ through the \texttt{dedn} interface. The interface returns the complete derivative for that bead, and only its root rank contributes to the average across beads obtained using \texttt{MPI\_Allreduce}. This average determines the potentiostat force $\mu_{\mathrm{e}}-P^{-1}\sum_{b=1}^{P}\partial U(\bm{R}^{(b)},N_{\mathrm{e}})/\partial N_{\mathrm{e}}$. Electronic propagation is inserted into the inherited force and coordinate updates: the electronic velocity and coordinate are updated on the first bead and broadcast from its root rank to all beads using \texttt{MPI\_Bcast} after each update. During thermostat propagation, the electronic velocity is scaled using the mean friction of the mode thermostats, and its kinetic-energy contribution is included in the driving term of each chain, reusing the same NHC propagator as \texttt{FixPIMDNVT}. Both \texttt{uvt} and \texttt{pimd/uvt} support restart files that preserve the electronic and thermostat states. The following example activates TP-PIMD with $\mu_{\mathrm{e}}=-3.0$\,eV, a potentiostat damping time of $0.05$\,ps, and an initial excess electron number of $0.648702$, specified by \texttt{mu}, \texttt{Udamp}, and \texttt{ne}, respectively. The compute \texttt{dEdN} supplies each bead's derivative using the evolving shared electron number, as discussed in Subsection~\ref{sec:2-2-2}:
\begin{grayverb}[nobreak=false]
\begin{Verbatim}
fix 1 all pimd/uvt method nmpimd temp 300 &
    Tdamp 0.05 tchain 3 mu -3.0 Udamp 0.05 &
    ne 0.648702 dedn c_dEdN
\end{Verbatim}
\end{grayverb}

\subsection{General electronic-response interface}
\label{sec:2-2-2}

The TP-Classical MD and TP-PIMD formulations both require a scalar electronic response. For TP-Classical MD, this response is $W_{\mathrm{e}}$, as defined in Eq.~\eqref{eq:5}. For TP-PIMD, the required response is the bead average $\overline{W}_{\mathrm{e}}$, as defined in Eq.~\eqref{eq:bead-averaged-response}. As illustrated in FIG.~\ref{fig:1}(c), both \texttt{FixUVT} and \texttt{FixPIMDUVT} classes call their \texttt{evaluate\_dedn} function during the \texttt{post\_force} stage to obtain the electron-number derivative through the common \texttt{dedn} interface. On the user side, \texttt{fix uvt} and \texttt{fix pimd/uvt} accept this derivative through the \texttt{dedn} keyword. The input can reference an equal-style \texttt{variable} (\texttt{v\_name}), an output from a \texttt{compute} (\texttt{c\_ID}), or an output from a \texttt{fix} (\texttt{f\_ID}). An equal-style \texttt{variable} evaluates a user-defined expression to a scalar value, whereas a \texttt{compute} or \texttt{fix} is a LAMMPS object whose output can be passed directly to the integrator. This common input convention supports analytical expressions or custom implementations of the electronic response, providing flexibility in its evaluation without modifying the propagation algorithm. In our examples, the analytical model supplies the derivative through the custom \texttt{fix uvt/toy/coupled} style, and the DP model uses the \texttt{compute deepmd/fparam/dedn} style implemented in DeePMD-kit,\cite{Wang_ComputPhysCommun_2018_v228_p178,Zeng_JChemPhys_2023_v159_p054801,Zeng_JChemTheoryComput_2025_v21_p4375} which uses a central finite difference to calculate $\partial U/\partial N_{\mathrm{e}}$.

In simulations distributed over multiple CPU cores, all MPI ranks within a LAMMPS partition evaluate the \texttt{dedn} source, which must return the same total derivative on every rank for the classical system or the corresponding PIMD bead. Any required summation over spatial domains is handled by the source; for example, \texttt{compute deepmd/fparam/dedn} uses \texttt{MPI\_Allreduce} to sum the domain contributions. The integrator therefore receives the complete derivative without requiring an additional sum within the partition.

\begin{figure*}[!t]
\centering
\includegraphics[width=0.9\linewidth,height=0.72\textheight,keepaspectratio]{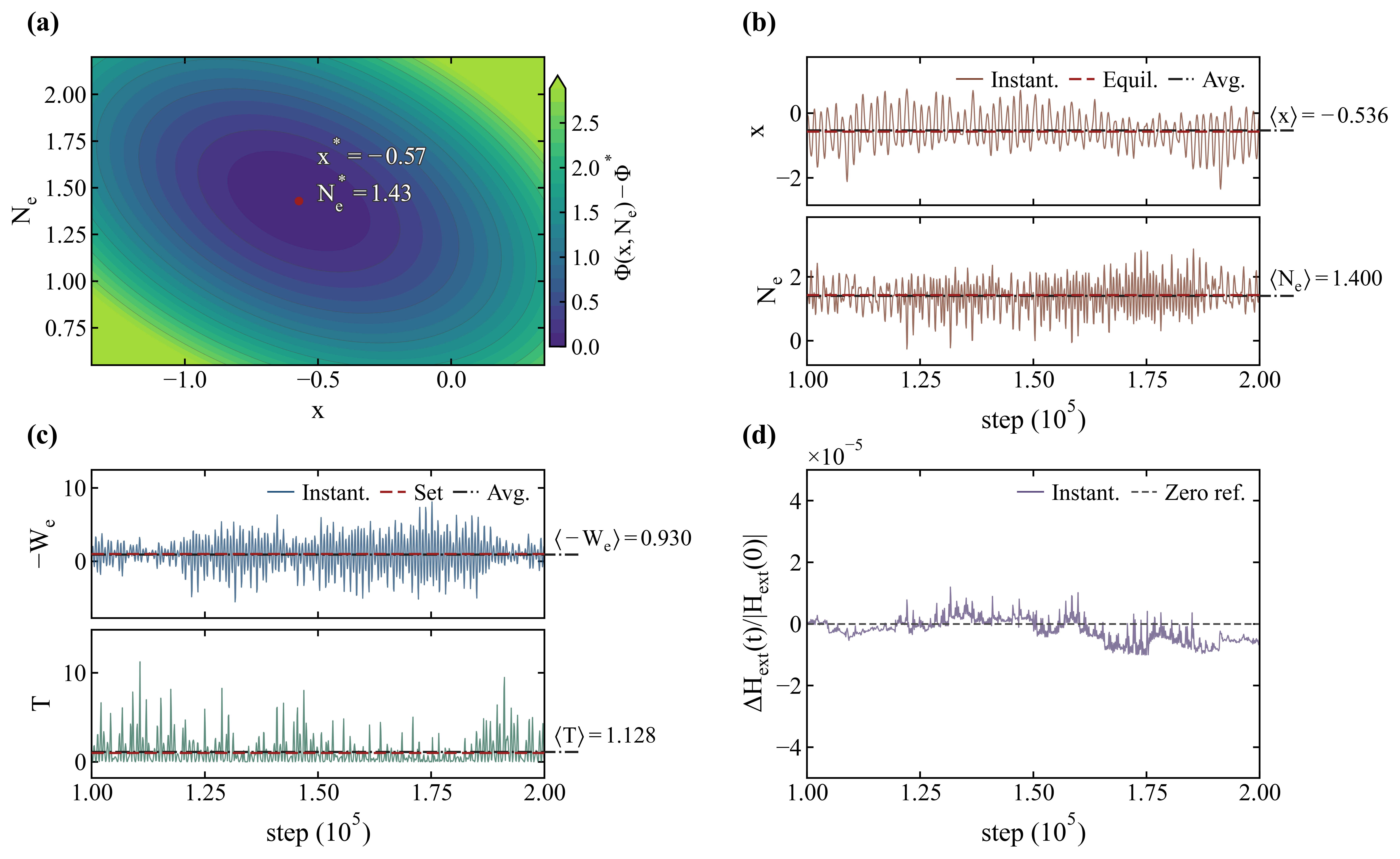}
\caption{Validation of the TP-Classical MD implementation using an analytical model. (a) Grand potential surface \(\itPhi\left( x,N_{\mathrm{e}} \right) - \itPhi^{*}\) at \(\mu_{\mathrm{e}} = 1.0\), with the minimum marked by a red circle. (b) Time evolution of \(x\) and \(N_{\mathrm{e}}\), compared with their equilibrium values obtained from grand-potential minimization. Solid curves denote instantaneous values, black dash-dotted lines denote time averages, and red dashed lines indicate the equilibrium values. (c) Instantaneous values of \({- W}_{\mathrm{e}}\) and \(T\), fluctuating around their corresponding set values. The line styles have the same meanings as in (b), with the red dashed line indicating the set values. (d) Relative fluctuation of the extended Hamiltonian \(\Delta H_{\text{ext}}(t)\text{/}\left| H_{\text{ext}}(0) \right|\). The gray dashed line indicates the zero reference. The first $10^{5}$ steps are discarded for equilibration.}
\label{fig:2}
\end{figure*}

\section{Examples}
\label{sec:3}

We present three examples to illustrate the usage and validate the implementation of the methods described in Sec.~\ref{sec:2-2}. We first examine TP-Classical MD using an analytical model and a machine learning model of an electrochemical interface. We then compare NHC and PILE\_L thermostatting for PIMD simulations of liquid water. Finally, we test TP-PIMD using the analytical model and the electrochemical interface.

\subsection{TP-Classical MD for an Analytical Model}
\label{sec:3-1}
\label{sec:classical-analytical}

We first validate the TP-Classical MD implementation using an analytical model that explicitly couples a representative particle coordinate \(x\) to the electron number \(N_{\mathrm{e}}\). The potential energy is defined as

\begin{equation}
\label{eq:16}
U\left( x,N_{\mathrm{e}} \right) = \frac{1}{2}k_{x}x^{2} + \frac{1}{2}k_{\mathrm{e}}\left( N_{\mathrm{e}} - N_{0} \right)^{2} + gxN_{\mathrm{e}},
\end{equation}

where \(k_{x}\) and \(k_{\mathrm{e}}\) are the corresponding harmonic coefficients, \(N_{0}\) is the reference electron number, and \(g\) controls the strength of the bilinear coupling. At a prescribed electrochemical potential \(\mu_{\mathrm{e}}\), the grand potential follows Eq.~\eqref{eq:2}. All quantities are expressed in reduced units. Using \(k_{x} = 5.0\), \(k_{\mathrm{e}} = 5.0\), \(g = 2.0\), \(N_{0} = 1.0\), and \(\mu_{\mathrm{e}} = 1.0\), with the temperature set to \(T = 1.0\), minimization of grand potential \(\itPhi(x,N_{\mathrm{e}})\) gives \(x^{*} = - 0.57\) and \(N_{\mathrm{e}}^{*} = 1.43\), as marked in FIG. 2(a). The analytical expressions for these equilibrium values and their derivation are provided in Sec.~S5 of the SM.

The model was propagated according to Eq.~\eqref{eq:4}. As shown in FIG. 2(b), both \(x\) and \(N_{\mathrm{e}}\) exhibit stationary thermal fluctuations around their analytical equilibrium values, with averages in close agreement with \(x^{*}\) and \(N_{\mathrm{e}}^{*}\). The constant-potential condition was independently assessed through \(- W_{\mathrm{e}}\). As shown in FIG. 2(c), \(- W_{\mathrm{e}}\) fluctuates around the imposed \(\mu_{\mathrm{e}} = 1.0\). The instantaneous temperature fluctuates around the set value \(T = 1.0\). Finally, the relative extended-Hamiltonian fluctuation remains bounded without secular growth over the trajectory {[}FIG. 2(d){]}, confirming the numerical stability of the integration. Hereafter, \(H_{\text{ext}}\) denotes the corresponding extended Hamiltonian of the simulation method.

\begin{figure*}[!t]
\centering
\includegraphics[width=0.8\linewidth,height=0.72\textheight,keepaspectratio]{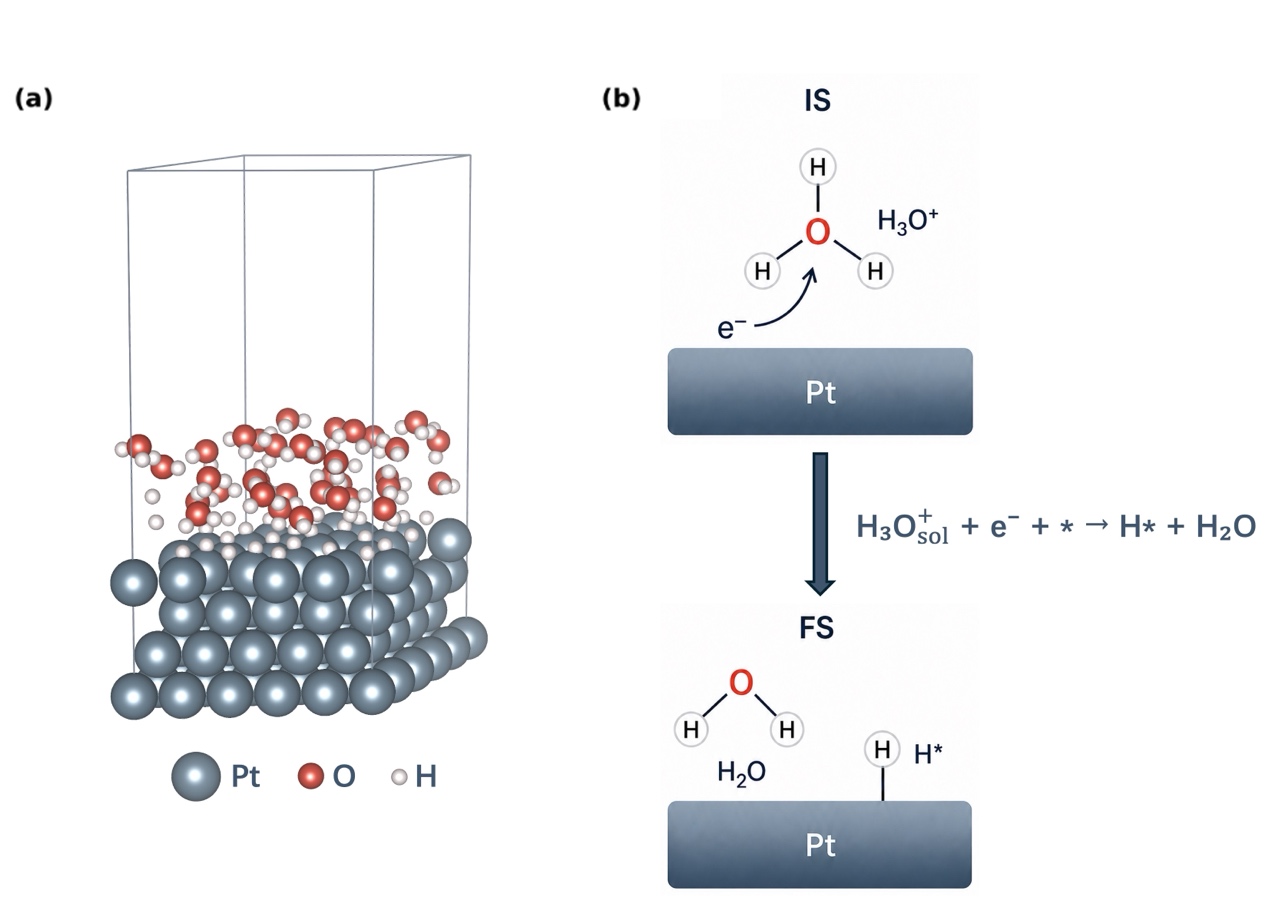}
\caption{Pt(111)--water interface model and Volmer reaction scheme. (a) Simulation cell containing a Pt(111) slab and an explicit interfacial water layer. (b) Schematic representation of the Volmer step from the solvated-proton reactant state to adsorbed hydrogen on the Pt surface.}
\label{fig:3}
\end{figure*}

\subsection{TP-Classical MD for a DP Interface Model}
\label{sec:classical-dp}

We next test the TP-Classical MD implementation with a realistic machine learning electrochemical model. The Volmer step at a Pt(111)--water interface was described using an electron-number-dependent Deep Potential (DP-\(N_{\mathrm{e}}\)) model\cite{Zhang2018Deep} that provides the potential energy, atomic forces, and electronic response.\cite{Fu2026ElectrochemistryEnhanced,Sun2025Probing} The interfacial model and the corresponding Volmer process are illustrated in FIG. 3. Additional details of the simulation cell and machine learning potential are provided in Sec. S2 of the SM.

\begin{figure*}[!t]
\centering
\includegraphics[width=0.9\linewidth,height=0.72\textheight,keepaspectratio]{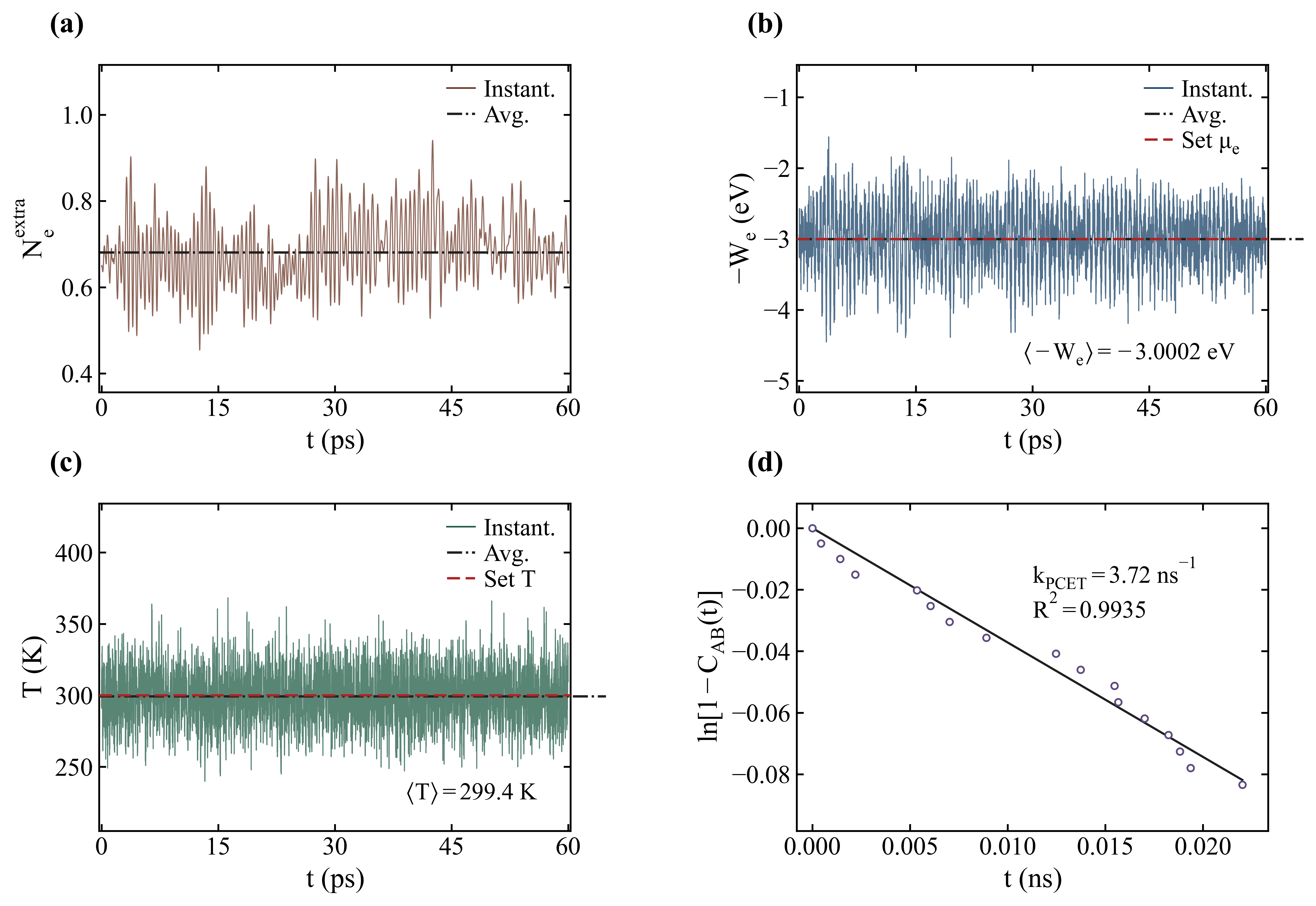}
\caption{TP-Classical MD validation and Volmer reaction-rate analysis using the machine learning model. (a) Time evolution of \(N_{\mathrm{e}}^{\text{extra}}\). The solid curve denotes the instantaneous value, and the black dash-dotted line denotes the time average. (b) Instantaneous \({- W}_{\mathrm{e}}\) fluctuations around the set electrochemical potential \(\mu_{\mathrm{e}} = - 3.0\,\text{eV}\) vs. vacuum. The line styles are the same as in (a), with the red dashed line indicating the set \(\mu_{\mathrm{e}}\). (c) Instantaneous temperature fluctuations around the set temperature \(T = 300\,\text{K}\). The line styles are the same as in (a), with the red dashed line indicating the set \(T\). (d) Side--side correlation analysis for reaction-rate extraction, plotted as \(\ln\left\lbrack 1 - C_{\mathrm{AB}}(t) \right\rbrack\) versus time at \(\mu_{\mathrm{e}} = - 1.8\,\text{eV}\) vs. vacuum, yielding \(k_{\text{PCET}} = 3.72\,\text{n}\text{s}^{-1}\).}
\label{fig:4}
\end{figure*}

The machine learning-based TP-Classical MD was first examined at \(T = 300\,\text{K}\) and \(\mu_{\mathrm{e}} = - 3.0\,\text{eV}\) relative to the vacuum level, corresponding to an electrode potential of -1.44 V vs. the Standard Hydrogen Electrode (SHE), using the absolute potential of the SHE (approximately 4.44\,V vs. vacuum).\cite{TrasattiThe} Following our previous JCTC paper,\cite{Fu2026ElectrochemistryEnhanced} we retain the convention that $N_{\mathrm{e}}$ denotes the total electron number in the theoretical formulation, while the DP interface examples use the excess electron number $N_{\mathrm{e}}^{\mathrm{extra}}=N_{\mathrm{e}}-N_{\mathrm{e}}^{\mathrm{neutral}}$ relative to the neutral reference state. For the present interface model containing 100 Pt, 36 O, and 97 H atoms, the neutral reference contains $N_{\mathrm{e}}^{\mathrm{neutral}}=8185$ electrons, including core electrons. The DP model and the integrator use this excess electron number as the propagated electronic coordinate; the constant reference shift leaves the electron-number derivative and equations of motion unchanged. The thermostat and potentiostat damping times were both 0.05\,ps, and the initial excess electron number was $N_{\mathrm{e}}^{\mathrm{extra}}=0.648702$. As shown in FIG. 4(a), \(N_{\mathrm{e}}^{\text{extra}}\) exhibits stationary fluctuations around a stable mean. The corresponding \({- W}_{\mathrm{e}}\) fluctuates around the imposed value {[}FIG. 4(b){]}. The instantaneous temperature remains centered around \(300\,\text{K}\) {[}FIG. 4(c){]}. These results show that the electron-number-dependent machine learning model can be coupled to the TP-Classical MD propagator through the common electronic-response interface while retaining the expected constant-potential sampling behavior.

Having established stable constant-potential sampling with the electron-number-dependent machine learning model, we next applied the TP-Classical MD framework to the kinetic analysis of the Volmer reaction at \(\mu_{\mathrm{e}} = - 1.8\,\text{eV}\) vs. vacuum. The corresponding proton-transfer process is illustrated in FIG. 3(b). We define a reactant region (A), corresponding to the solvated-proton initial state (IS), and a product region (B), corresponding to the final state (FS) with hydrogen adsorbed on the Pt surface. The detailed geometric criteria used to assign molecular configurations to the IS and FS, together with the persistence criterion used to exclude transient fluctuations, are given in Sec. S3 of the SM.

The transition kinetics were quantified using the side--side time-correlation function,\cite{Fu2026ElectrochemistryEnhanced,Frenkel2023Chapter}

\begin{equation}
\label{eq:17}
C_{\mathrm{AB}}(t) = \frac{\left\langle h_{\mathrm{A}}\left\lbrack x(0) \right\rbrack h_{\mathrm{B}}\left\lbrack x(t) \right\rbrack \right\rangle}{\left\langle h_{\mathrm{A}}\left\lbrack x(0) \right\rbrack \right\rangle}.
\end{equation}

Here, \(x(t)\) represents the phase-space coordinates of the system at time \(t\), while \(h_{\mathrm{A}}\) and \(h_{\mathrm{B}}\) are indicator functions for the reactant (A) and product (B) regions, taking a value of unity when the system lies within the corresponding region and zero otherwise. Accordingly, \(C_{\mathrm{AB}}(t)\) represents the conditional probability of finding the system in the product region at time \(t\), given that it was initially in the reactant region. The correlation function is evaluated statistically from an ensemble of MD trajectories based on their time-dependent state assignments.

For a two-state process, the relaxation time \(\tau_{\mathrm{R}}\) is related to the forward and reverse rate constants through

\begin{equation}
\label{eq:18}
{\tau_{\mathrm{R}}}^{- 1}{= k}_{\mathrm{A}\rightarrow\mathrm{B}} + k_{\mathrm{B}\rightarrow\mathrm{A}}.
\end{equation}

Under the present reaction conditions, reverse transitions from the product region to the reactant region are assumed to be negligible, such that \(k_{\mathrm{B}\rightarrow\mathrm{A}} = 0\). The forward rate constant is therefore denoted by \(k_{\text{PCET}}{= k}_{\mathrm{A}\rightarrow\mathrm{B}}\), and the side--side correlation function reduces to

\begin{equation}
\label{eq:19}
C_{\mathrm{AB}}(t) = 1 - e^{- k_{\mathrm{PCET}}t}.
\end{equation}

Accordingly,

\begin{equation}
\label{eq:20}
\ln\left\lbrack 1 - C_{\mathrm{AB}}(t) \right\rbrack = - k_{\text{PCET}}t,
\end{equation}

so that \(k_{\text{PCET}}\) can be obtained from the negative slope of a linear fit of \(\ln\left\lbrack 1 - C_{\mathrm{AB}}(t) \right\rbrack\) as a function of time, with the fit constrained to pass through the origin. As shown in FIG. 4(d), the resulting linear regression yields \(k_{\text{PCET}} = 3.72\,\text{n}\text{s}^{-1}\). The upper limit of the fitting interval was selected based on the maximum coefficient of determination \(R^{2}\), obtained from linear regression; further details provided in Sec. S3 of the SM. The results demonstrate the framework's capability for dynamic calculations of electrochemical reaction rates.

\begin{figure*}[!t]
\centering
\includegraphics[width=0.9\linewidth,height=0.72\textheight,keepaspectratio]{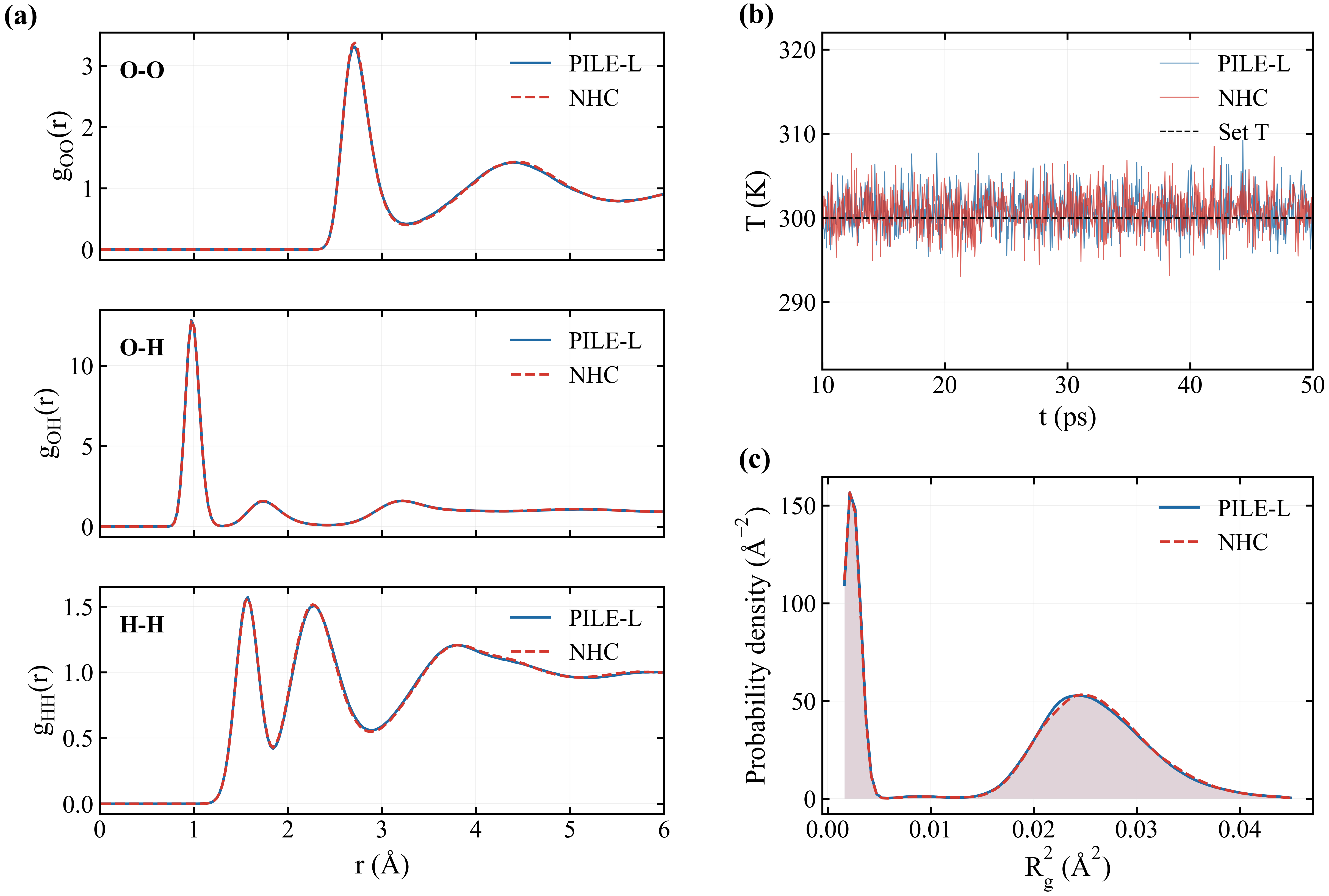}
\caption{PIMD sampling validation for liquid water. Results from 32-bead simulations using the PILE\_L and NHC thermostats. (a) Bead-averaged radial distribution functions \(g_{\text{OO}}(r)\), \(g_{\text{OH}}(r)\), and \(g_{\text{HH}}(r)\). (b) Instantaneous temperature \(T\) with the set temperature indicated by the black dashed line. (c) Distributions of \(R_{\mathrm{g}}^{2}\), accumulated over all water atoms and sampled configurations. Blue and red curves denote the PILE\_L and NHC results, respectively.}
\label{fig:5}
\end{figure*}

\subsection{PIMD with an NHC Thermostat for DP Water}
\label{sec:3-2}

To validate the newly implemented NHC thermostat\cite{Tuckerman1993Efficient,Martyna1992NosHoover} used in the PIMD propagation framework, we compared its equilibrium sampling with that obtained using the established PILE\_L thermostat\cite{Ceriotti2010Efficient} for bulk liquid water under a canonical ensemble ($NVT$) condition. The PILE\_L reference was obtained using the existing \texttt{fix pimd/langevin} implementation in LAMMPS, which employs a Langevin thermostat of the ring polymer modes, whereas the present implementation uses a Nosé--Hoover chain thermostat. A system containing 128 water molecules with 32 beads was simulated using a previously developed DP model for liquid water.\cite{Li2026Fix,Li2025Assessment} Simulations were performed at 300 K with a timestep of 0.5 fs for $1.0\times10^5$ steps. The PILE\_L and our NHC PIMD simulations used the same code for $NVE$ propagation of the ring polymer and only differ in thermostatting.

The equilibrium structural properties were first compared using bead-averaged radial distribution functions. As shown in FIG. 5(a), the \(g_{\text{OO}}(r)\), \(g_{\text{OH}}(r)\), and \(g_{\text{HH}}(r)\) obtained with the PILE\_L and NHC thermostats closely agree over the sampled distance range.

The sampling behavior of the two thermostats was further examined using the instantaneous temperature and the distribution of the squared atomic ring polymer radius of gyration. For each atom \emph{i}, the latter is defined as

\begin{equation}
\label{eq:21}
R_{\mathrm{g},i}^{2} = \frac{1}{P}\sum_{b = 1}^{P}\left| \bm{r}_{i}^{(b)} - {\overline{\bm{r}}}_{i} \right|^{2},
\end{equation}

where \({\overline{\bm{r}}}_{i} = \frac{1}{P}\sum_{b = 1}^{P}\bm{r}_{i}^{(b)}\) is the centroid of the corresponding atomic ring polymer. The \(R_{\mathrm{g}}^{2}\) distribution was accumulated over all water atoms and sampled configurations. As shown in FIG. 5(b), both thermostat schemes maintain temperature fluctuations centered around the set value throughout the trajectories. The \(R_{\mathrm{g}}^{2}\) distributions obtained with the two thermostats closely overlap over the sampled range {[}FIG. 5(c){]}. Together with the agreement in the radial distribution functions, these results show that the NHC implementation yields equilibrium sampling consistent with PILE\_L under the present simulation conditions.

\begin{figure*}[!t]
\centering
\includegraphics[width=0.9\linewidth,height=0.72\textheight,keepaspectratio]{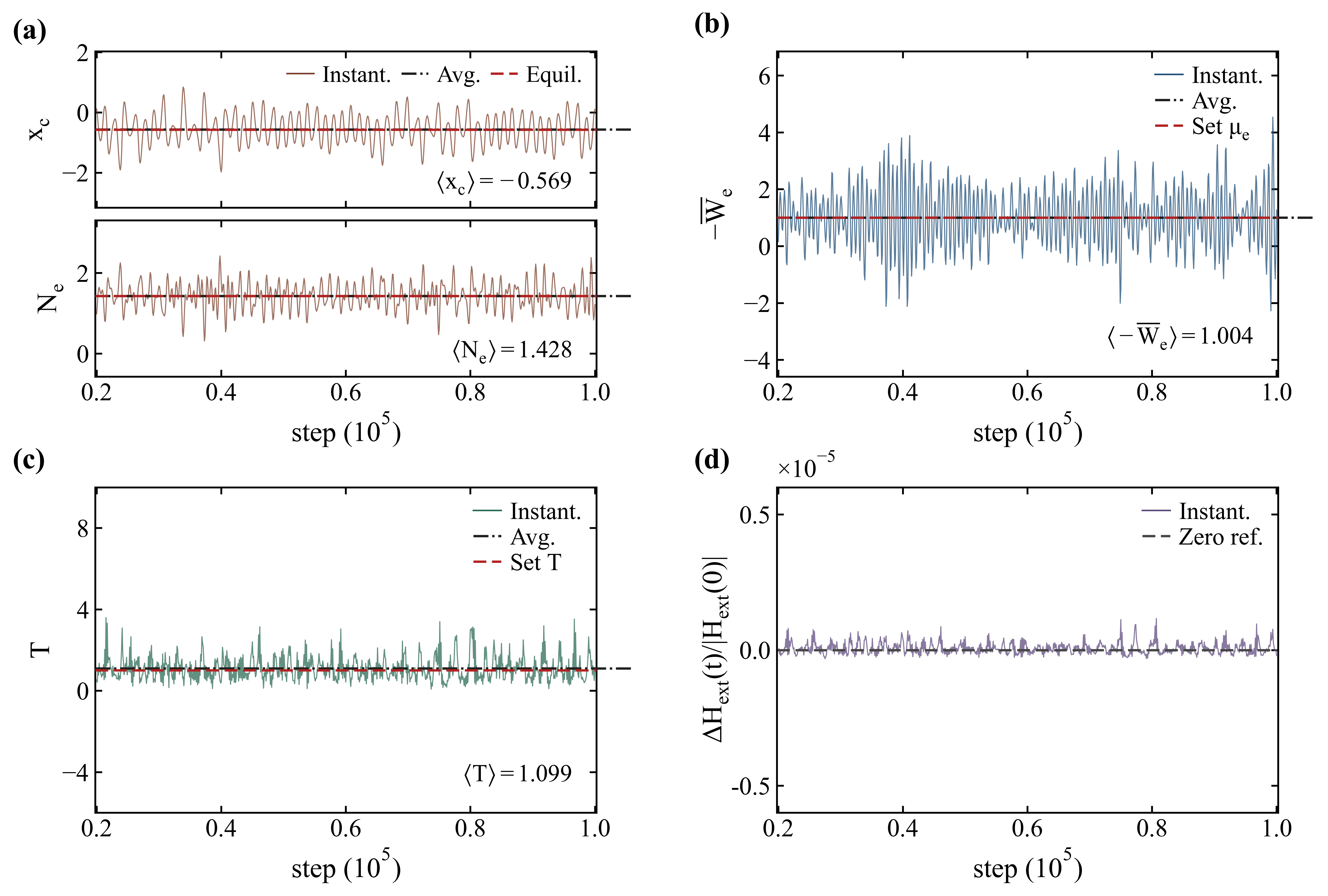}
\caption{Validation of the TP-PIMD implementation using the analytical coupled model with \(P = 8\) beads. (a) Time evolution of \(x_{\mathrm{c}}\) and \(N_{\mathrm{e}}\). Solid curves denote instantaneous values, black dash-dotted lines denote time averages, and red dashed lines indicate the equilibrium values obtained from grand-potential minimization. (b) Instantaneous \(- {\overline{W}}_{\mathrm{e}}\). The line styles have the same meanings as in (a), with the red dashed line indicating the set \(\mu_{\mathrm{e}} = 1.0\). (c) Instantaneous \(T\). The line styles have the same meanings as in (a), with the red dashed line indicating the set \(T = 1.0\). (d) Relative extended-Hamiltonian fluctuation \(\Delta H_{\text{ext}}(t)\text{/}\left| H_{\text{ext}}(0) \right|\). The gray dashed line indicates the zero reference.}
\label{fig:6}
\end{figure*}

\subsection{TP-PIMD for an Analytical Model}
\label{sec:3-3}

Having validated the TP-Classical MD and thermostatted-PIMD implementations separately, we next assess their combined operation within the TP-PIMD framework. The analytical model introduced in Sec.~\ref{sec:3-1} was first examined using \(P = 8\) and the same reduced-unit parameters. As shown in FIG. 6(a), both the centroid particle coordinate \(x_{\mathrm{c}} = \frac{1}{P}\sum_{b = 1}^{P}x^{(b)}\) and \(N_{\mathrm{e}}\) exhibit stationary fluctuations around their equilibrium values. \({- \overline{W}}_{\mathrm{e}}\) fluctuates around the set value \(\mu_{\mathrm{e}} = 1.0\), with its average satisfying Eq.~\eqref{eq:15} {[}FIG. 6(b){]}. The instantaneous temperature fluctuates around the set value \(T = 1.0\) {[}FIG. 6(c){]}, while the relative extended-Hamiltonian fluctuation remains bounded without systematic growth over the trajectory {[}FIG. 6(d){]}. Additional simulations with \(P = 1\) and \(P = 16\) show that \(\left\langle x_{\mathrm{c}} \right\rangle\), \(\left\langle N_{\mathrm{e}} \right\rangle\) and \(\left\langle - {\overline{W}}_{\mathrm{e}} \right\rangle\) remain consistent with their expected equilibrium values within the estimated block standard errors. The corresponding ring polymer radius of gyration \(\left\langle R_{\mathrm{g}}^{2} \right\rangle\) further confirms the finite bead delocalization introduced by the path integral representation for \(P > 1\). The detailed results are provided in Sec. S4 of the SM.

\begin{figure*}[!t]
\centering
\includegraphics[width=0.9\linewidth,height=0.72\textheight,keepaspectratio]{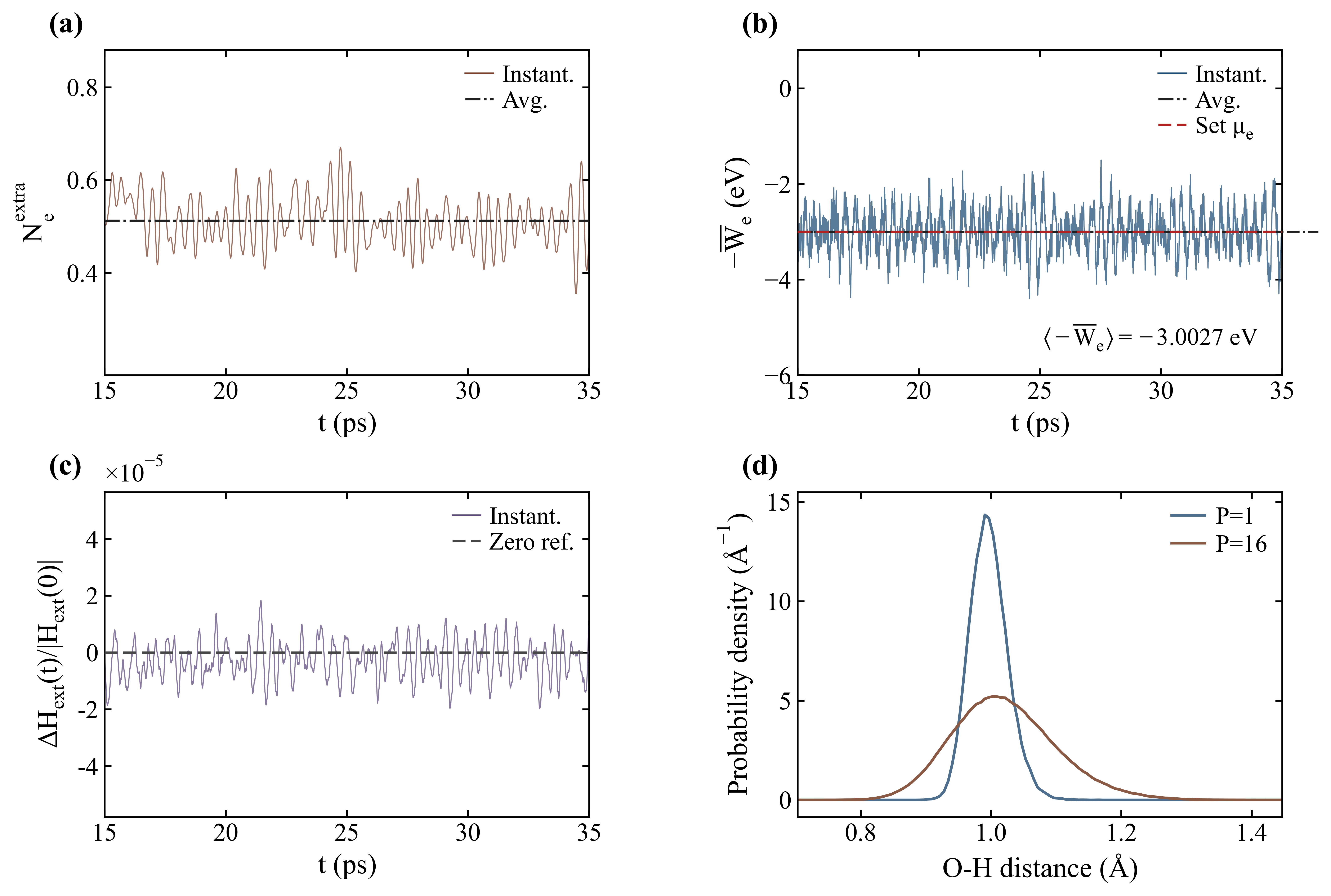}
\caption{TP-PIMD simulation of a machine learning electrochemical interface with \(P = 16\) beads. Representative \(P = 16\) results at \(T = 300\,\text{K}\) and \(\mu_{\mathrm{e}} = - 3.0\,\text{eV}\) vs. vacuum, showing (a) Time evolution of \(N_{\mathrm{e}}^{\text{extra}}\). Solid curves denote instantaneous values, black dash-dotted lines denote time averages. (b) Instantaneous \({- \overline{W}}_{\mathrm{e}}\). The line styles have the same meanings as in (a), with the red dashed line indicating \(\mu_{\mathrm{e}} = - 3.0\,\text{eV}\) vs. vacuum. (c) Relative extended-Hamiltonian fluctuation \(\Delta H_{\text{ext}}(t)\text{/}\left| H_{\text{ext}}(0) \right|\), with the gray dashed line indicating the zero reference. (d) Normalized nearest-neighbor O--H distance distributions for water hydrogens obtained from \(P = 1\) and \(P = 16\) TP-PIMD simulations under otherwise identical conditions.}
\label{fig:7}
\end{figure*}

\subsection{TP-PIMD for a DP Interface Model}
\label{sec:pimd-dp}

The TP-PIMD implementation was further tested using the Pt(111)--water interface introduced in Sec.~\ref{sec:classical-dp} (\textbf{FIG. 3}). The simulation was performed using \(P = 16\) at \(T = 300\,\text{K}\) and \(\mu_{\mathrm{e}} = - 3.0\,\text{eV}\) vs. vacuum. The thermostat and potentiostat damping times were both 0.05\,ps, and the initial excess electron number was $N_{\mathrm{e}}^{\mathrm{extra}}=0.648702$, following the convention described in Sec.~\ref{sec:classical-dp}. As shown in FIG. 7(a), the excess electron number \(N_{\mathrm{e}}^{\text{extra}}\) exhibits stationary fluctuations around a stable mean. The instantaneous \({- \overline{W}}_{\mathrm{e}}\) fluctuates around the imposed \(\mu_{\mathrm{e}}\) {[}FIG. 7(b){]}, consistent with the constant-potential condition under ring polymer sampling. The relative extended-Hamiltonian fluctuation remains bounded without systematic growth over the trajectory {[}FIG. 7(c){]}, indicating stable numerical integration. To examine nuclear quantum sampling at the interface, we compared the normalized nearest-neighbor O--H distance distribution of water hydrogens from \(P = 16\) and \(P = 1\) TP-PIMD simulations under otherwise identical conditions. For the \(P = 16\) TP-PIMD simulation, the distribution was accumulated over all beads and sampled configurations. As shown in FIG. 7(d), the \(P = 16\) distribution is substantially broader than the corresponding \(P = 1\) distribution and exhibits enhanced probability at longer O--H distances, consistent with enhanced nuclear quantum fluctuations along the O--H coordinate. These results demonstrate that the implemented TP-PIMD framework captures nuclear quantum fluctuations in realistic constant-potential electrochemical interfaces.

\section{Conclusions}
\label{sec:4}

In this work, we have developed a LAMMPS-native framework that integrates TP-Classical MD and TP-PIMD within a common simulation infrastructure. The framework propagates a shared electronic DOF and provides a general electronic-response interface for evaluating the electronic response using different computational models, including analytical expressions, finite-difference evaluations, and machine learning-based models. The implementation establishes a unified PIMD hierarchy within LAMMPS while preserving compatibility with standard MD workflows, including parallel execution and restart functionality. Validation across analytical models, electron-number-dependent machine learning models, and PIMD sampling benchmarks demonstrates the reliability and versatility of the framework. This development provides a general and reusable computational framework for studying electrochemical systems under constant-potential conditions with nuclear quantum effects and machine learning-based interaction models. The framework establishes a foundation for future investigations of complex electrochemical interfaces and reactive processes.

\section*{Supplementary Material}
The SM provides the relation between physical-temperature and scaled-temperature representations; details of the Pt(111)--water interface model and electron-number-dependent machine learning potential; the procedure for calculating Volmer reaction rates; additional validation of TP-PIMD sampling; the analytical equilibrium solution of the coupled model; and the derivation of the equilibrium constant-potential condition.

\begin{acknowledgments}
Y.L. thanks Axel Gomez and Axel Kohlmeyer for discussions on the code design. L.F. and S.X. gratefully acknowledge funding support from the National Natural Science Foundation of China (grant no. 92470114, no. 52273223), Ministry of Science and Technology of the People's Republic of China (grant no. 2021YFB3800303), DP Technology Corporation (grant no. 2021110016001141), School of Materials Science and Engineering at Peking University, and the AI for Science Institute, Beijing (AISI). The computing resource of this work was provided by the Bohrium Cloud Platform (https://bohrium.dp.tech), which was supported by DP Technology.

\end{acknowledgments}

\section*{Author Declarations}

\subsection*{Conflict of Interest}
The authors have no conflicts to disclose.

\subsection*{Author Contributions}
\textbf{Li Fu:} Conceptualization (equal); Software (equal); Investigation (equal); Writing -- original draft (equal); Writing -- review \& editing (equal). \textbf{Yifan Li:} Conceptualization (equal); Software (equal); Investigation (equal); Writing -- original draft (equal); Writing -- review \& editing (equal). \textbf{Shenzhen Xu:} Conceptualization (equal); Investigation (equal); Writing -- original draft (equal); Writing -- review \& editing (equal); Funding acquisition.

\section*{Data Availability}
The code developed in this work is openly available through Pull Request 5107, submitted to the official LAMMPS repository: \url{https://github.com/lammps/lammps/pull/5107}. The simulation inputs, model files, and processed data supporting this study are openly available at \url{https://github.com/Lily200202/LAMMPS_uvt_MD_PIMD_Examples}.\cite{LAMMPSUVTExamples2026}

\clearpage
\setcounter{section}{0}
\setcounter{figure}{0}
\setcounter{equation}{0}
\setcounter{table}{0}
\renewcommand{\thesection}{S\arabic{section}}
\renewcommand{\thefigure}{S\arabic{figure}}
\renewcommand{\theequation}{S\arabic{equation}}
\renewcommand{\thetable}{S\arabic{table}}

\begin{center}
{\large\bfseries Supplementary Material}\par\medskip
{\large fix uvt and fix pimd/uvt: A Unified LAMMPS Framework for Constant-Potential Constant-Temperature Molecular Dynamics}\par\medskip
Li Fu, Yifan Li, and Shenzhen Xu
\end{center}
\bigskip

\section{Relation between physical-temperature and scaled-temperature representations}
\label{sec:si-1}

The thermostatted-potentiostatted path integral molecular dynamics (TP-PIMD) formulation presented in the main text Sec.~II~B is written in the physical-temperature representation, whereas the LAMMPS implementation employs the equivalent scaled-temperature representation used in the underlying PIMD framework.\cite{Thompson2022LAMMPS,Li2026Fix} This section summarizes the correspondence between these two representations.

The ring polymer propagation Hamiltonian in the physical-temperature representation can be uniformly scaled by the number of beads:

\begin{equation}
\label{eq:si-1}
H_{\text{scaled}} = PH_{\text{phys}}.
\end{equation}

Accordingly, the inverse temperature associated with the scaled representation is

\begin{equation}
\label{eq:si-2}
\beta_{\text{scaled}} = \frac{\beta}{P},
\end{equation}

which preserves the same equilibrium configurational distribution.

In the physical-temperature representation, the ring polymer frequency used in Sec.~II~B is

\begin{equation}
\label{eq:si-3}
\omega_{P} = \sqrt{P}\text{/}(\beta\hslash).
\end{equation}

The corresponding frequency in the scaled-temperature representation used by LAMMPS becomes

\begin{equation}
\label{eq:si-4}
\omega_{P}^{\text{LAMMPS}} = \frac{P}{\beta\hslash}.
\end{equation}

Consistently, the thermostat temperature of the scaled ring polymer system is

\begin{equation}
\label{eq:si-5}
T_{\text{scaled}} = PT.
\end{equation}

Therefore, the temperature reported by the scaled-temperature ring polymer system corresponds to P times the physical temperature defined in Sec.~II~B, and the physical temperature is obtained by dividing the LAMMPS ring polymer temperature by the bead number P.

\section{Pt(111)--water interface model and electron-number-dependent machine learning potential}
\label{sec:si-2}

The Pt(111)--water interface model employed in the ~thermostatted-potentiostatted classical molecular dynamics (TP-Classical MD) and TP-PIMD simulations was adopted from our previous electrochemical interface study.\cite{Fu2026ElectrochemistryEnhanced} The electrode was represented by a 5\(\times\)5 Pt(111) slab consisting of four atomic layers, with the bottom two Pt layers constrained during MD simulations. The electrolyte region contained a water bilayer with 36 explicit water molecules and one solvated proton initially represented as a hydronium ion H\textsubscript{3}O\textsuperscript{+}. The Pt surface was prepared with a pre-adsorbed hydrogen coverage of one monolayer (1 ML), leaving one vacant adsorption site for the Volmer reaction. The simulation cell dimensions were 14.06\(\times\)14.06\(\times\)30.89 Å\cite{Fu2026ElectrochemistryEnhanced}, and periodic boundary conditions were applied.

The interatomic interactions were described using the electron-number-dependent Deep Potential model (DP-\(N_{\mathrm{e}}\)), which was developed to describe electrochemical systems with variable electron numbers.\cite{Fu2026ElectrochemistryEnhanced,Sun2025Probing} In this framework, the total electron number is treated as an explicit variable of the potential-energy surface, allowing the potential energy to be expressed as \(U\left( \bm{R},N_{\mathrm{e}} \right)\). The resulting dependence on \(N_{\mathrm{e}}\) enables evaluation of the electronic response required for constant-potential MD simulations.

The DP-\(N_{\mathrm{e}}\) model used here was trained using density functional theory (DFT) reference data generated with the ABACUS package\cite{Chen2010Systematically,Li2016LargeScale}. The exchange--correlation interaction was described using the Perdew--Burke--Ernzerhof (PBE)\cite{Perdew1996Generalized} generalized-gradient approximation together with the Grimme D3 dispersion correction\cite{Grimme2010A}. The accuracy of the trained DP-\(N_{\mathrm{e}}\) model was evaluated against independent DFT reference calculations using testing configurations sampled from both classical and quantum trajectories, showing good agreement for the predicted energies and atomic forces; detailed benchmark results have been reported previously.\cite{Fu2026ElectrochemistryEnhanced} The validated model was then used to provide the potential energy, atomic forces, and electronic response required by the constant-potential framework.

\begin{figure*}[!t]
\centering
\includegraphics[width=0.85\textwidth]{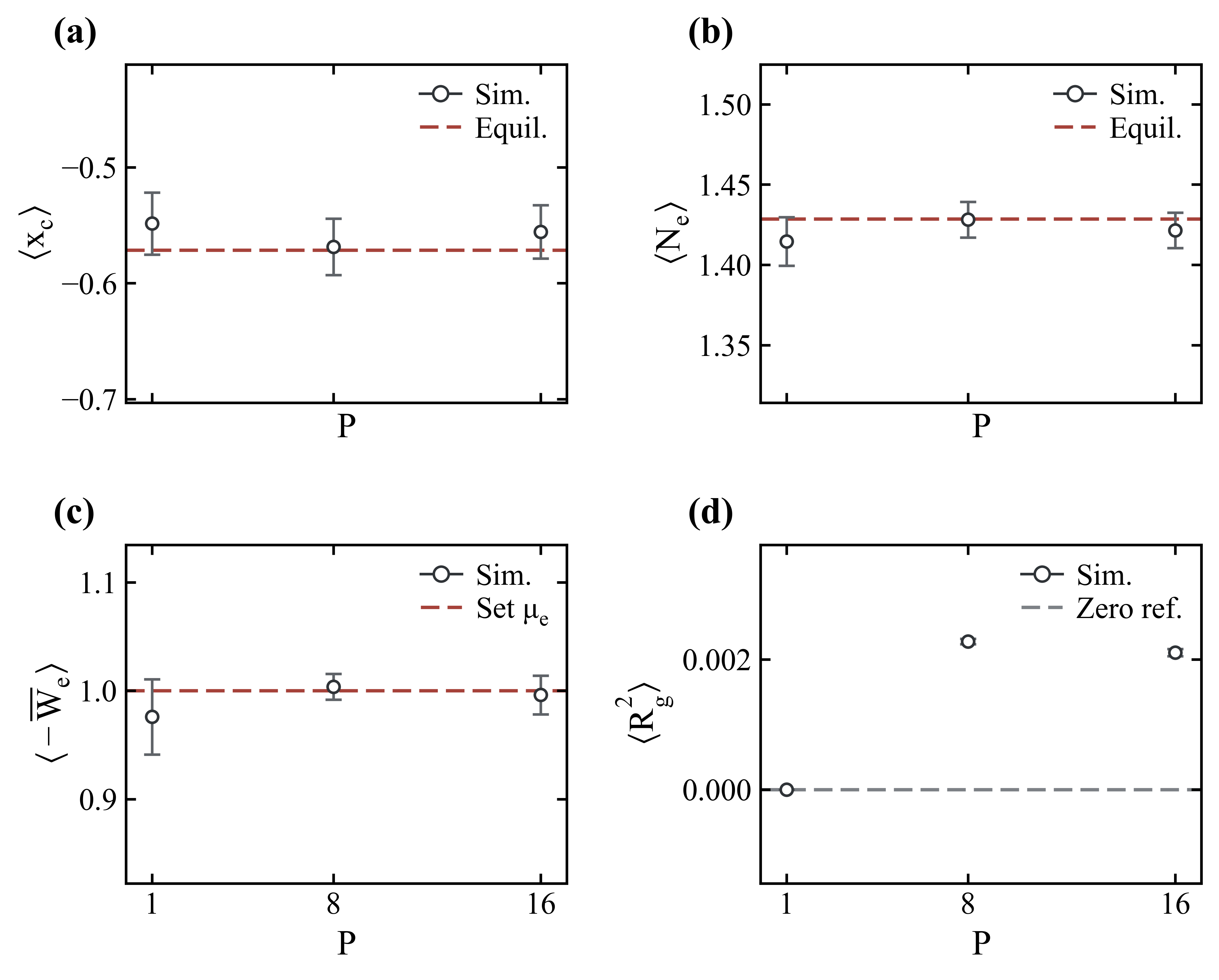}
\caption{Bead-number convergence and sampling consistency of TP-PIMD simulation using an analytical model. (a) Centroid particle coordinate \(\left\langle x_{\mathrm{c}} \right\rangle\), with the red dashed line indicating the equilibrium value \(x^{*} = - 0.57\). (b) Shared electron-number variable \(\left\langle N_{\mathrm{e}} \right\rangle\), with the red dashed line indicating the equilibrium value \(N_{\mathrm{e}}^{*} = 1.43\). (c) Negative bead-averaged electronic response \(\left\langle {- \overline{W}}_{\mathrm{e}} \right\rangle\), with the red dashed line indicating the imposed electrochemical potential \(\mu_{\mathrm{e}} = 1.0\). (d) Ring polymer radius of gyration \(\left\langle R_{\mathrm{g}}^{2} \right\rangle\) for \(P = 1\), 8, and 16. The gray dashed line in (d) denotes the classical-limit reference \(R_{\mathrm{g}}^{2} = 0\). Error bars denote block standard errors over the production window.}
\label{fig:S1}
\end{figure*}

\section{Details of dynamic calculations of Volmer reaction rates}
\label{sec:si-3}

The Volmer reaction rate constant was evaluated using a side--side time-correlation function analysis.\cite{Frenkel2023Chapter} The initial-state (IS) and final-state (FS) definitions and reaction-rate extraction procedure were adopted from our previous study.\cite{Fu2026ElectrochemistryEnhanced} A proton transfer event was identified using geometric criteria based on the vertical position of the transferred hydrogen and the nearest oxygen--hydrogen distance. Specifically, the transferred hydrogen was considered to reach the final state when \(z_{\text{H}} < z_{\text{Pt}} + 1.7\ \mathring{\mathrm{A}}\), where \(z_{\text{Pt}}\) denotes the average vertical position of the top-layer Pt atoms. In addition, the distance between the transferred hydrogen and its nearest oxygen atom was required to satisfy \(r_{\text{O-H}} > 1.6\ \mathring{\mathrm{A}}\). The combination of these two criteria defines the final state corresponding to hydrogen adsorption at the Pt surface. To avoid counting transient structural fluctuations as reaction events, a persistence criterion was applied: a trajectory was considered to enter the final state only when the final-state condition was satisfied for more than half of 100 consecutive simulation steps. Please refer to our group's earlier paper for more details of the model setup.\cite{Fu2026ElectrochemistryEnhanced} The fitting interval was determined by scanning different upper time limits \(t^{*}\) and selecting the interval that maximized the coefficient of determination \(R^{2}\).\cite{Blumer2024ShortTime}

\section{Additional validation of TP-PIMD}
\label{sec:si-4}

To further assess the bead-number dependence of TP-PIMD, additional simulations were performed using the analytical model introduced in Sec.~IV~A of the main text. Simulations with \(P = 1\), 8, and 16 beads were carried out under identical thermodynamic conditions. The convergence of the path integral sampling was evaluated using the centroid particle coordinate, \(x_{\mathrm{c}} = \frac{1}{P}\sum_{b = 1}^{P}x^{(b)}\), the shared electron-number variable \(N_{\mathrm{e}}\), and the bead-averaged electronic response \({\overline{W}}_{\mathrm{e}} = \frac{1}{P}\sum_{b = 1}^{P}W_{\mathrm{e}}^{(b)}\), where \(W_{\mathrm{e}}^{(b)} = - \frac{\partial U\left( \bm{R}^{(b)},\ \ N_{\mathrm{e}} \right)}{\partial N_{\mathrm{e}}}\). The ring polymer delocalization was quantified using the radius of gyration \(R_{\mathrm{g}}^{2} = \frac{1}{P}\sum_{b = 1}^{P}\left( x^{(b)} - x_{\mathrm{c}} \right)^{2}\). For the imposed electrochemical potential \(\mu_{\mathrm{e}} = 1.0\), the equilibrium values of the coupled model are \(x^{*} = - 0.57\) and \(N_{\mathrm{e}}^{*} = 1.43\). The temperature was set to \(T = 1.0\).

As shown in FIG. S1, the averaged quantities obtained from simulations with different bead numbers remain consistent with the corresponding equilibrium values within the estimated statistical uncertainties. The \(P = 1\) simulation recovers the classical constant-potential limit, while increasing the bead number introduces finite ring polymer delocalization associated with the path integral representation of nuclear quantum effects while preserving the expected equilibrium values of \(x_{\mathrm{c}}\), \(N_{\mathrm{e}}\), and \({- \overline{W}}_{\mathrm{e}}\).

In particular, the bead-averaged electronic response satisfies the constant-potential condition \(\left\langle {- \overline{W}}_{\mathrm{e}} \right\rangle \approx \mu_{\mathrm{e}}\), for all examined bead numbers, confirming that the shared electron-number DOF remains correctly coupled to the electronic reservoir in the path integral formulation. The statistical uncertainties were estimated using block averaging. The production trajectory was divided into 20 equal blocks. Error bars denote block standard errors, computed as the sample standard deviation of the 20 block averages divided by \(\sqrt{20}\).

\section{Equilibrium solution of the analytical model}
\label{sec:si-analytical-equilibrium}

For the coupled harmonic model introduced in Sec.~IV~A of the main text, the grand potential is
\begin{equation}
\begin{aligned}
\Phi(x,N_{\mathrm{e}}) ={}& \frac{1}{2}k_x x^2
+ \frac{1}{2}k_{\mathrm{e}}(N_{\mathrm{e}}-N_0)^2 \\
&+ gxN_{\mathrm{e}}-\mu_{\mathrm{e}}N_{\mathrm{e}}.
\end{aligned}
\end{equation}
Its stationary point satisfies $\partial\Phi/\partial x=0$ and $\partial\Phi/\partial N_{\mathrm{e}}=0$, giving
\begin{equation}
\begin{aligned}
k_x x^{*}+gN_{\mathrm{e}}^{*} &= 0, \\
k_{\mathrm{e}}(N_{\mathrm{e}}^{*}-N_0)+gx^{*}-\mu_{\mathrm{e}} &= 0.
\end{aligned}
\end{equation}
Solving these equations yields
\begin{equation}
\label{eq:si-analytical-equilibrium}
\begin{aligned}
x^{*} &= -\frac{g(k_{\mathrm{e}}N_0+\mu_{\mathrm{e}})}{k_x k_{\mathrm{e}}-g^2}, \\
N_{\mathrm{e}}^{*} &= \frac{k_x(k_{\mathrm{e}}N_0+\mu_{\mathrm{e}})}{k_x k_{\mathrm{e}}-g^2}.
\end{aligned}
\end{equation}
For $k_x>0$ and $k_x k_{\mathrm{e}}-g^2>0$, the grand potential is strictly convex, so this stationary point is its unique global minimum. With $k_x=k_{\mathrm{e}}=5.0$, $g=2.0$, $N_0=1.0$, and $\mu_{\mathrm{e}}=1.0$, the solution is $x^{*}=-4/7\approx-0.57$ and $N_{\mathrm{e}}^{*}=10/7\approx1.43$, providing the equilibrium reference values used in the main text.

\section{Derivation of the equilibrium constant-potential condition}
\label{sec:si-constant-potential}

In the continuous electron-number formulation of Sec.~II~A of the main text, the normalized equilibrium distribution is
\begin{equation}
\label{eq:si-gc-density}
\begin{aligned}
\rho_{\mathrm{cl}}(\bm R,\bm P,N_{\mathrm{e}})
={}& Z^{-1}\exp\bigl\{-\beta\bigl[K(\bm P) \\
&+U(\bm R,N_{\mathrm{e}})-\mu_{\mathrm{e}}N_{\mathrm{e}}\bigr]\bigr\},
\end{aligned}
\end{equation}
where $K(\bm P)=\sum_i\bm p_i^2/(2m_i)$ and $Z$ is the normalization constant. At fixed $T$ and $\mu_{\mathrm{e}}$, differentiation with respect to $N_{\mathrm{e}}$ gives
\begin{equation}
\label{eq:si-density-derivative}
\frac{\partial\rho_{\mathrm{cl}}}{\partial N_{\mathrm{e}}}
=-\beta\left(\frac{\partial U}{\partial N_{\mathrm{e}}}-\mu_{\mathrm{e}}\right)\rho_{\mathrm{cl}}.
\end{equation}
Let $N_{\min}$ and $N_{\max}$ denote the boundaries of the electron-number domain, which may be infinite. We assume that the distribution is differentiable and normalizable and that its boundary values vanish. For fixed $\bm R$ and $\bm P$, the fundamental theorem of calculus then gives
\begin{equation}
\label{eq:si-boundary-term}
\begin{aligned}
\int_{N_{\min}}^{N_{\max}}\mathrm dN_{\mathrm{e}}\,
\frac{\partial\rho_{\mathrm{cl}}}{\partial N_{\mathrm{e}}}
&=\left[\rho_{\mathrm{cl}}\right]_{N_{\min}}^{N_{\max}} \\
&=0.
\end{aligned}
\end{equation}
Integrating also over the particle positions and momenta, and substituting Eq.~\eqref{eq:si-density-derivative}, yields
\begin{equation}
\label{eq:si-average-derivative}
\begin{aligned}
0&=\int\mathrm d\Gamma\,
\frac{\partial\rho_{\mathrm{cl}}}{\partial N_{\mathrm{e}}} \\
&=-\beta\int\mathrm d\Gamma\,
\left(\frac{\partial U}{\partial N_{\mathrm{e}}}-\mu_{\mathrm{e}}\right)\rho_{\mathrm{cl}} \\
&=-\beta\left(\left\langle\frac{\partial U}{\partial N_{\mathrm{e}}}\right\rangle-\mu_{\mathrm{e}}\right),
\end{aligned}
\end{equation}
where $\mathrm d\Gamma=\mathrm d\bm R\,\mathrm d\bm P\,\mathrm dN_{\mathrm{e}}$. The final equality follows from the definition of an ensemble average, $\langle A\rangle=\int\mathrm d\Gamma\,A\rho_{\mathrm{cl}}$, and normalization, $\int\mathrm d\Gamma\,\rho_{\mathrm{cl}}=1$. Since $W_{\mathrm{e}}=-\partial U/\partial N_{\mathrm{e}}$, we obtain
\begin{equation}
\label{eq:si-constant-potential}
\left\langle-W_{\mathrm{e}}\right\rangle=\mu_{\mathrm{e}}.
\end{equation}
Thus, the electronic chemical potential equals the reservoir value on average, while its instantaneous value can fluctuate. This result relies on the vanishing boundary term in Eq.~\eqref{eq:si-boundary-term}.

\bibliographystyle{aipnum4-2}
\bibliography{references}

\end{document}